\documentclass[a4paper,11pt]{article}
\pdfoutput=1
\usepackage{jheppub}
\usepackage[T1]{fontenc}
\usepackage{graphicx}
\usepackage{amssymb}
\usepackage{amsmath}
\usepackage{booktabs}
\usepackage{multirow}
\usepackage{bbm}
\usepackage{CJKutf8}
\usepackage{listings}
\usepackage[dvipsnames]{xcolor}
\lstdefinestyle{mystyle}{
  language=Mathematica,
  backgroundcolor=\color{white},
  basicstyle=\ttfamily\small,
  keywordstyle=\color{blue}\bfseries,
  commentstyle=\color{TealBlue},
  stringstyle=\color{gray},
  frame=tb, 
  numbers=left,
  numberstyle=\tiny\color{gray},
  stepnumber=1,
  numbersep=5pt,
  showstringspaces=false,
  tabsize=2,
  breaklines=true,
  breakatwhitespace=false,
  captionpos=b
}
\allowdisplaybreaks

\title{\boldmath Construction of general $N$-body lattice operators with arbitrary momenta}

\author[a,b]{Haobo~Yan~(\begin{CJK*}{UTF8}{gbsn}燕浩波\end{CJK*}),}
\author[a,c,d]{Chuan~Liu~(\begin{CJK*}{UTF8}{gbsn}刘川\end{CJK*}),}
\author[e,f]{Liuming~Liu~(\begin{CJK*}{UTF8}{gbsn}刘柳明\end{CJK*})}
\author[g]{and Yu~Meng~(\begin{CJK*}{UTF8}{gbsn}孟雨\end{CJK*})}

\affiliation[a]{School of Physics, Peking University,\\Beijing 100871, China}
\affiliation[b]{Helmholtz-Institut f\"ur Strahlen- und Kernphysik (Theorie) and Bethe Center for Theoretical Physics, Universit\"at Bonn,\\53115 Bonn, Germany}
\affiliation[c]{Center for High Energy Physics, Peking University,\\Beijing 100871, China}
\affiliation[d]{Collaborative Innovation Center of Quantum Matter,\\Beijing 100871, China}
\affiliation[e]{Institute of Modern Physics, Chinese Academy of Sciences,\\Lanzhou 730000, China}
\affiliation[f]{University of Chinese Academy of Sciences,\\Beijing 100049, China}
\affiliation[g]{School of Physics, Zhengzhou University,\\Zhengzhou 450001, China}

\emailAdd{haobo@stu.pku.edu.cn}
\emailAdd{liuchuan@pku.edu.cn}
\emailAdd{liuming@impcas.ac.cn}
\emailAdd{yu\_meng@zzu.edu.cn}

\abstract{We present a systematic method for constructing lattice QCD operators for systems of an arbitrary number of particles with arbitrary momentum, spin, and internal quantum numbers. Explicit constructions are provided for one-, two-, three-, and four-hadron operators, covering all irreducible representations of the relevant lattice symmetry groups in both rest and moving frames. The construction procedure has been implemented in the open-source package \texttt{OpTion} (Operator construcTion), available at \url{https://github.com/wittscien/OpTion}. The paper and the package are designed to serve as a practical and extensible dictionary for future lattice QCD studies, as lattice calculations advance towards increasingly complex hadronic systems.}

\begin{document} 
\maketitle
\flushbottom

\section{Introduction}
\label{sec:intro}

Over the past few decades, both experimental and theoretical efforts have witnessed a flourishing development in hadron spectroscopy. Since the establishment of the Standard Model, a large number of resonance particles have been discovered, as well as exotic hadronic states that cannot be described by the conventional quark model. For these particles - both the well-known resonances, like $\rho$, $\omega$, and the exotic states, named by $X$,$Y$,$Z$ particles - a unified understanding of their properties has been challenging due to the non-perturbative nature of the strong interaction. Although phenomenological studies have made significant progress in these areas, see~\cite{Chen:2016qju, Guo:2017jvc, Olsen:2017bmm, Brambilla:2019esw, Chen:2022asf} for recent reviews, many questions remain unresolved. Lattice QCD, as a first-principles computational method in quantum field theory, has evolved into an indispensable and crucial tool in current spectroscopic research~\cite{Briceno:2017max, Hansen:2019nir, Mai:2021lwb, Mai:2022eur}.

In typical lattice QCD computations, particles and the corresponding fields are
defined on a four-dimensional discretized hypercubic spacetime lattice. Physical observables are then estimated in Monte Carlo simulations from Euclidean space correlation functions. One of the four dimensions is picked out as the so-called temporal direction to facilitate the signal of the correlation function. For example, in typical spectrum computations, one is more concerned with two-point functions of the form: $\langle \mathcal{O}_i(t_1)\mathcal{O}_j(t_2)\rangle$ of various operators at time-slice $t$: $\{\mathcal{O}_i(t): i=1,\cdots N_{\mathrm{op}}\}$ that are built from basic lattice fields, i.e. the quark, anti-quark and possibly the gauge fields. The correlation function can be sent through a standard generalized eigen problem (GEVP) routine, and towers of finite-volume energy levels can be extracted. With a larger set of operators, more information can be explored on the energy region of interest. In the case of lattice computations of more complex physical quantities, three- or four-point functions are also needed. For example, isolating contributions from $N\pi$ excited states in the nucleon electric polarizabilities demands the inclusion of interpolating operators with appropriate transformation properties under the cubic group~\cite{Wang:2023omf}. In any case, the construction of the appropriate hadron operators $\{\mathcal{O}_i(t): i=1,\cdots N_{\mathrm{op}}\}$ is a starting point and becomes common practice in almost all lattice QCD computations.

In the lattice setup described above, the hadron operators built on a particular time slice $t$ normally need to carry certain quantum numbers respected by QCD. Another symmetry respected by the lattice is the finite translation symmetry in spatial directions characterized by the center of mass (CM) three-momenta $\vec{P}$. Therefore, translation and three-dimensional rotation symmetry in continuum QCD is broken down to, taking into account possible Lorentz boosts, the subgroups of the three-dimensional space group with underlying cubic symmetry. The corresponding finite-volume spectra are also subduced into the irreducible representations (irreps) of the corresponding little groups. In this paper, we consider the CM three-momenta $\vec{P} = [0,0,0]$, $[0,0,n]$, $[0,n,n]$, $[n,n,n]$, $[n,m,0]$, $[n,n,m]$, $[n,m,p]$ in the unit of $\frac{2\pi}{L}$ for non-zero integers $n \neq m \neq p$. These choices span all distinct momentum classes allowed on a cubic lattice, up to permutations and sign flips. They are sufficient to cover the relevant irreps of the lattice symmetry group. Various $N$-particle hadron operators are constructed which transform within a particular irreps. These $N$-hadron operators can be utilized in the computation of correlation functions in any lattice study.

The construction of hadron operators on a lattice has been studied over the years by many lattice groups~\cite{Basak:2005aq, Basak:2005ir, Basak:2007kj, Chen:2014afa, Prelovsek:2016iyo, CLQCD:2019npr, Detmold:2024ifm, Morningstar:2013bda, Wallace:2015pxa, Lyu:2022tsd}. In this paper, we try to synthesize previous results and establish a unified and consistent convention. These results are generalized and are now incorporated into a \texttt{Mathematica} package 
called \texttt{OpTion}, for Operator construcTion, which can be accessed through Github~\cite{github}. The paper also aims to serve as a practical reference and dictionary for future lattice studies.

The organization of this paper is as follows. In Section~\ref{sec:groups}, we summarize various little groups and their irreps that arise from the reduced symmetry on a cubic lattice. Section~\ref{sec:one} presents our methodology for constructing operators, including a list of one-particle operators involving up to one covariant derivative. Based upon these, Sections~\ref{sec:two} and~\ref{sec:int} describe the construction of general $N$-particle operators and their decomposition into relevant internal quantum numbers. A brief tutorial for the \texttt{OpTion} code, along with an explicit list of two-particle operators, is provided in the appendix~\ref{sec:tutorials}, \ref{sec:append_list_mm}, \ref{sec:append_list_bb}, and \ref{sec:append_list_mb}.

\section{Lattice groups}
\label{sec:groups}
As the CM momentum $\vec{P}$ is varied among $\Omega = \{[0,0,0]$, $[0,0,n]$, $[0,n,n]$, $[n,n,n]$, $[n,m,0]$, $[n,n,m]$, $[n,m,p]\}$, the symmetry of the lattice is gradually reduced and eventually breaks down to a trivial group, i.e., all spatial symmetry operations are broken. Furthermore, for half-integer total spin, the double-cover version of the group must be utilized. In this work, we consider all irreps of the lattice groups. The corresponding symmetry groups and links to the irreps are detailed in Table~\ref{tab:group}. As a summary, the dictionary of the one-hadron and two-hadron operators is tabulated, and the corresponding links are collected in Tab.~\ref{tab:group}.

\begin{table}[htbp]
\centering
\caption{The corresponding symmetry groups and their double covers for $\vec{P} \in \Omega$, the corresponding irrep tables, along with the summary of the one-hadron and two-hadron operators, are listed. Here, $n \neq m \neq p$ are non-zero integers, and groups with a superscript $D$ indicate the double cover.}
\addtolength{\tabcolsep}{-5pt}
\begin{tabular}{cccccccc}
\toprule
$\vec{P}$ & $[0,0,0]$ & $[0,0,n]$ & $[0,n,n]$ & $[n,n,n]$ & $[n,m,0]$ & $[n,n,m]$ & $[n,m,p]$ \\
\midrule
group & $O_h$ & $C_{4v}$ & $C_{2v}$ & $C_{3v}$ & $C_{2}$ & $C_{2}$ & $C_{1}$ \\
irreps & Sec.~\ref{sec:oh} & Tab.~\ref{tab:c4v} & Tab.~\ref{tab:c2v} & Tab.~\ref{tab:c3v} & Tab.~\ref{tab:c2} & Tab.~\ref{tab:c2} & Tab.~\ref{tab:c1} \\
one-hadron & Tab.~\ref{tab:one-000} & Tab.~\ref{tab:one-00n} & Tab.~\ref{tab:one-0nn} & Tab.~\ref{tab:one-nnn} & Tab.~\ref{tab:one-nm0} & Tab.~\ref{tab:one-nnm} & Tab.~\ref{tab:one-nmp} \\
meson-meson & Tab.~\ref{tab:mm-000-A1+}-\ref{tab:mm-000-T2-} & Tab.~\ref{tab:mm-00n-A1}-\ref{tab:mm-00n-E} & Tab.~\ref{tab:mm-0nn-A1}-\ref{tab:mm-0nn-B2} & - & - & - & - \\
baryon-baryon & Tab.~\ref{tab:bb-000} & Tab.~\ref{tab:bb-00n} & Tab.~\ref{tab:bb-0nn} & Tab.~\ref{tab:bb-nnn} & Tab.~\ref{tab:bb-nm0} & Tab.~\ref{tab:bb-nnm} & - \\
\midrule
double cover & $O_h^D$ & $C_{4v}^D$ & $C_{2v}^D$ & $C_{3v}^D$ & $C_{2}^D$ & $C_{2}^D$ & $C_{1}^D$ \\
irreps & Sec.~\ref{sec:oh} & Tab.~\ref{tab:c4vd} & Tab.~\ref{tab:c2vd} & Tab.~\ref{tab:c3vd} & Tab.~\ref{tab:c2d} & Tab.~\ref{tab:c2d} & Tab.~\ref{tab:c1d} \\
meson-baryon & Tab.~\ref{tab:mb-000-G1}-\ref{tab:mb-000-H-} & Tab.~\ref{tab:mb-00n} & Tab.~\ref{tab:mb-0nn} & - & - & - & - \\
\bottomrule
\end{tabular}
\addtolength{\tabcolsep}{5pt}
\label{tab:group}
\end{table}

\subsection{Cubic groups}
\label{sec:oh}
In the rest frame, the largest lattice symmetry group is $O_h = O \otimes \{E, I\}$, where $O$ is the proper octahedral group consisting of $24$ elements, organized into $5$ conjugacy classes: $I$, $C_3$, $C_4$, $C_2^{\prime}$, and $C_2$. The group elements, corresponding to the fundamental representation $T_1$, can be parameterized as (see, e.g., Ref.~\cite{Bernard:2008ax}):
\begin{equation}
    (R_i)_{\alpha \beta}=\exp (-i \vec{n}_i \cdot \vec{J} \omega_i)_{\alpha \beta}=\cos \omega_i \delta_{\alpha \beta}+\left(1-\cos \omega_i\right) n_{i\alpha} n_{i\beta}-\sin \omega_i \epsilon_{\alpha \beta \gamma} n_{i\gamma},
\end{equation}
where $(J_\gamma)_{\alpha \beta} = -i \epsilon_{\alpha \beta \gamma}$ are the group generators, $\vec{n}$ is the rotation axis, and $\omega$ is the rotation angle. For systems involving half-integer spin, the double cover of the octahedral group, denoted $O_h^D$, must be used. The group contains $48$ elements falling into 8 conjugacy classes: $I$, $C_6$, $C_3$, $C_8^{\prime}$, $C_8$, $C_4^{\prime}$, and $C_4$. The explicit group elements, their conjugacy classes, and how each element is generated are summarized in Table~\ref{tab:oh}. This table is equivalent to the one in
Ref.~\cite{Bernard:2008ax}. The ordering for the $O$ group remains unchanged, while for the $O_h^D$ group, the first $24$ elements correspond to those of the original $O$ group, with the remaining $24$ elements representing the original elements supplemented by an additional $\pm 2\pi$ rotation.

\begin{table}
\centering
\caption{Elements of the (double cover) cubic group and their classification. The first $24$ elements belong to the $O_h$ group, while the remaining $24$ are the additional elements in the $O_h^D$ group. The last column indicates how each element can be generated from $C_{4x}$ and $C_{4y}$ rotations.}
\begin{tabular}{ccccc}
\toprule
Class & i & $\vec{n}$ & $\omega$ & \\
\midrule
$I(I/J)$ & $1 / 25$ & any & $0 /+2 \pi$ & - \\
\midrule
\multirow{8}{*}{$8 C_3\left(8 C_6 / 8 C_3\right)$} & $2 / 26$ & $\frac{1}{\sqrt{3}}(1,1,1)$ & $-\frac{2 \pi}{3} /+2 \pi$ & $C_{4 z} C_{4 y}$ \\
& $3 / 27$ & $\frac{1}{\sqrt{3}}(1,1,1)$ & $\frac{2 \pi}{3} /-2 \pi$ & $C_{4 y}^{-1} C_{4 z}^{-1}$ \\
& $4 / 28$ & $\frac{1}{\sqrt{3}}(-1,1,1)$ & $-\frac{2 \pi}{3} /+2 \pi$ & $C_{4 y} C_{4 z}$ \\
& $5 / 29$ & $\frac{1}{\sqrt{3}}(-1,1,1)$ & $\frac{2 \pi}{3} /-2 \pi$ & $C_{4 z}^{-1} C_{4 y}^{-1}$ \\
& $6 / 30$ & $\frac{1}{\sqrt{3}}(-1,-1,1)$ & $-\frac{2 \pi}{3} /+2 \pi$ & $C_{4 z} C_{4 y}^{-1}$ \\
& $7 / 31$ & $\frac{1}{\sqrt{3}}(-1,-1,1)$ & $\frac{2 \pi}{3} /-2 \pi$ & $C_{4 y} C_{4 z}^{-1}$ \\
& $8 / 32$ & $\frac{1}{\sqrt{3}}(1,-1,1)$ & $-\frac{2 \pi}{3} /+2 \pi$ & $C_{4 y}^{-1} C_{4 z}$ \\
& $9 / 33$ & $\frac{1}{\sqrt{3}}(1,-1,1)$ & $\frac{2 \pi}{3} /-2 \pi$ & $C_{4 z}^{-1} C_{4 y}$ \\
\midrule
\multirow{6}{*}{$6 C_4\left(6 C_8^{\prime} / 6 C_8\right)$} & $10 / 34$ & $(1,0,0)$ & $-\frac{\pi}{2} /+2 \pi$ & $C_{4 z} C_{4 y} C_{4 z}^{-1}$ \\
& $11 / 35$ & $(1,0,0)$ & $\frac{\pi}{2} /-2 \pi$ & $C_{4 z} C_{4 y}^{-1} C_{4 z}^{-1}$ \\
& $12 / 36$ & $(0,1,0)$ & $-\frac{\pi}{2} /+2 \pi$ & $C_{4 y}$ \\
& $13 / 37$ & $(0,1,0)$ & $\frac{\pi}{2} /-2 \pi$ & $C_{4 y}^{-1}$ \\
& $14 / 38$ & $(0,0,1)$ & $-\frac{\pi}{2} /+2 \pi$ & $C_{4 z}$ \\
& $15 / 39$ & $(0,0,1)$ & $\frac{\pi}{2} /-2 \pi$ & $C_{4 z}^{-1}$ \\
\midrule
\multirow{6}{*}{$6 C_2^{\prime}\left(12 C_4^{\prime}\right)$} & $16 / 40$ & $(0,1,1)$ & $-\pi /+2 \pi$ & $C_{4 y} C_{4 z} C_{4 y}$ \\
& $17 / 41$ & $(0,-1,1)$ & $-\pi /+2 \pi$ & $C_{4 y}^{-1} C_{4 z} C_{4 y}^{-1}$ \\
& $18 / 42$ & $(1,1,0)$ & $-\pi /+2 \pi$ & $C_{4 z} C_{4 y} C_{4 y}$ \\
& $19 / 43$ & $(1,-1,0)$ & $-\pi /+2 \pi$ & $C_{4 z}^{-1} C_{4 y}^{-1} C_{4 y}^{-1}$ \\
& $20 / 44$ & $(1,0,1)$ & $-\pi /+2 \pi$ & $C_{4 z} C_{4 z} C_{4 y}$ \\
& $21 / 45$ & $(-1,0,1)$ & $-\pi /+2 \pi$ & $C_{4 y} C_{4 z} C_{4 z}$ \\
\midrule
\multirow{3}{*}{$6 C_2\left(6 C_4\right)$} & $22 / 46$ & $(1,0,0)$ & $-\pi /+2 \pi$ & $C_{4 z} C_{4 z} C_{4 y} C_{4 y}$ \\
& $23 / 47$ & $(0,1,0)$ & $-\pi /+2 \pi$ & $C_{4 y} C_{4 y}$ \\
& $24 / 48$ & $(0,0,1)$ & $-\pi /+2 \pi$ & $C_{4 z} C_{4 z}$ \\
\bottomrule
\end{tabular}
\label{tab:oh}
\end{table}

The irreps of the $O^{(D)}$ group are compiled in Ref.~\cite{Bernard:2008ax} and are repeated here for completeness, along with additional details relevant for practical construction:

$A_1$: $\Gamma_{A_1} = 1$,

$A_2:$ $\Gamma_{A_2}=-1$ for classes $C_4$ and $C_2^{\prime}, \,\Gamma_{A_2}=1$ otherwise,

$E$:
\begin{alignat*}{2}
& \Gamma_{E}=1 \quad && \text{for} \quad i=1,22,23,24, \\
& \Gamma_{E}=\sigma_3 \quad && \text{for} \quad i=14,15,18,19, \\
& \Gamma_{E}=-\cos \frac{\pi}{3} \mathbf{1}+i \sin \frac{\pi}{3} \sigma_2 && \text{for} \quad i=2,5,6,9, \\
& \Gamma_{E}=-\cos \frac{\pi}{3} 1-i \sin \frac{\pi}{3} \sigma_2 && \text{for} \quad i=3,4,7,8, \\
& \Gamma_{E}=-\cos \frac{\pi}{3} \sigma_3-\sin \frac{\pi}{3} \sigma_1 && \text{for} \quad i=10,11,16,17, \\
& \Gamma_{E}=-\cos \frac{\pi}{3} \sigma_3+\sin \frac{\pi}{3} \sigma_1 \quad && \text{for} \quad i=12,13,20,21,
\end{alignat*}

$T_1$: $\Gamma_{T_1} = \exp (-i \vec{n}_i \cdot \vec{J} \omega_i)$,

$T_2$: the same as $\Gamma_{T_1}$, except for the change of sign for classes $C_4$ and $C_2^{\prime}$.

For the double cover $O^D$ of the cubic group $O$, irreps of $O$ are still present. These irreps are blind to the distinction between group elements that differ by a $2\pi$ rotation. In addition to these, $O^D$ contains three additional irreps that are unique to the double cover:

$G_1$: $\Gamma_{G_1}\equiv Y=\exp (-\frac{i}{2} \vec{n}_i \cdot \vec{\sigma} \omega_i)$,

$G_2$ : the same as $\Gamma_{G_1}$, except for the change of sign for classes $C_8, C_8^{\prime}$ and $C_4^{\prime}$,

$H$ : $\Gamma_{H}=\exp (-i \vec{n}_i \cdot \vec{J}^{\frac{3}{2}} \omega_i)$, where $\vec{J}^{\frac{3}{2}}$ are the generators for $SU(2)$ group for $J = 3 / 2$:

\begin{equation}
S_x =\begin{pmatrix}
0 & \tfrac{\sqrt{3}}{2} & 0 & 0 \\
\tfrac{\sqrt{3}}{2} & 0 & 1 & 0 \\
0 & 1 & 0 & \tfrac{\sqrt{3}}{2} \\
0 & 0 & \tfrac{\sqrt{3}}{2} & 0
\end{pmatrix},
S_y =\begin{pmatrix}
0 & -i \tfrac{\sqrt{3}}{2} & 0 & 0 \\
i \tfrac{\sqrt{3}}{2} & 0 & - i & 0 \\
0 &  i & 0 & -i \tfrac{\sqrt{3}}{2} \\
0 & 0 & i \tfrac{\sqrt{3}}{2} & 0
\end{pmatrix},
S_z =\begin{pmatrix}
\tfrac{3}{2} & 0 & 0 & 0 \\
0 & \frac{1}{2} & 0 & 0 \\
0 & 0 & -\tfrac{1}{2} & 0 \\
0 & 0 & 0 & -\tfrac{3}{2}
\end{pmatrix}.
\end{equation}

\subsection{Little groups}
In frames with non-zero total momentum, the symmetry is further reduced to the little groups that preserve that momentum under transformation. Specifically, the momentum configurations $[n,m,0]$ and $[n,n,m]$ reduce the symmetry to $C_2^{(D)}$, albeit with slight variations in their respective operations. To streamline the notation, the $X$ matrices introduced in Ref.~\cite{Gockeler:2012yj} are employed.
\begin{equation}
\begin{cases}
X_1=\mathbbm{1}, \\
X_2=-\frac{1}{2} \mathbbm{1}+i \frac{\sqrt{3}}{2} \sigma_2, \\
X_3=-\frac{1}{2} \mathbbm{1}-i \frac{\sqrt{3}}{2} \sigma_2, \\
X_4=-\frac{1}{2} \sigma_3-\frac{\sqrt{3}}{2} \sigma_1, \\
X_5=\sigma_3, \\
X_6=-\frac{1}{2} \sigma_3+\frac{\sqrt{3}}{2} \sigma_1, \\
X_7=i \frac{1}{\sqrt{2}}\left(\sigma_1+\sigma_2\right), \\
X_8=\frac{1}{\sqrt{2}}\left(\sigma_1-\sigma_2\right).
\end{cases}    
\end{equation}
We also denote $\omega = \operatorname{e}^{\frac{\pi i}{4}}$ and parity inversion $I$. Tables~\ref{tab:c4v} to table~\ref{tab:c1d} list the elements in each group. The specific table number is coded in Tab.~\ref{tab:oh}. For double cover groups, the element $R_{25} \equiv J$ denotes the $2\pi$ rotation.

\begin{table}
\centering
\caption{Elements and irreps for group $C_{4v}$.}
\begin{tabular}{cccccc}
\toprule
$g$ & $E$ & $\left\{R_{14}, R_{15}\right\}$ & $\left\{I R_{18}, I R_{19}\right\}$ & $\left\{I R_{22}, I R_{23}\right\}$ & $R_{24}$ \\
\midrule
$A_1$ & $1$ & $1$ & $1$ & $1$ & $1$ \\
$A_2$ & $1$ & $1$ & $-1$ & $-1$ & $1$ \\
$B_1$ & $1$ & $-1$ & $-1$ & $1$ & $1$ \\
$B_2$ & $1$ & $-1$ & $1$ & $-1$ & $1$ \\
$E$ & $2$ & $0$ & $0$ & $0$ & $-2$ \\
\midrule
$E$ & $X_1$ & $\left\{-X_7, X_7\right\}$ & $\left\{X_5,-X_5\right\}$ & $\left\{X_8,-X_8\right\}$ & $-X_1$ \\
\bottomrule
\end{tabular}
\label{tab:c4v}
\end{table}

\begin{table}
\scriptsize
\centering
\caption{Elements and irreps for group $C_{4v}^D$.}
\begin{tabular}{cccccccc}
\toprule
$g$ & $E$ & $\left\{R_{48}, R_{24}\right\}$ & $\left\{R_{15}, R_{14}\right\}$ & $\left\{R_{38}, R_{39}\right\}$ & $\left\{I R_{46}, I R_{47}, I R_{22}, I R_{23}\right\}$ & $\left\{I R_{42}, I R_{43}, I R_{18}, I R_{19}\right\}$ & $R_{25}$ \\
\midrule
$A_1$ & $1$ & $1$ & $1$ & $1$ & $1$ & $1$ & $1$ \\
$A_2$ & $1$ & $1$ & $1$ & $1$ & $-1$ & $-1$ & $1$ \\
$B_1$ & $1$ & $1$ & $-1$ & $-1$ & $1$ & $-1$ & $1$ \\
$B_2$ & $1$ & $1$ & $-1$ & $-1$ & $-1$ & $1$ & $1$ \\
$E$ & $2$ & $-2$ & $0$ & $0$ & $0$ & $0$ & $2$ \\
$G_1$ & $2$ & $0$ & $\sqrt{2}$ & $-\sqrt{2}$ & $0$ & $0$ & $-2$ \\
$G_2$ & $2$ & $0$ & $-\sqrt{2}$ & $\sqrt{2}$ & $0$ & $0$ & $-2$ \\
\midrule
$E$ & $1$ & $\{-\mathbbm{1},-\mathbbm{1}\}$ & $\left\{i \sigma_3,-i \sigma_3\right\}$ & $\left\{-i \sigma_3, i \sigma_3\right\}$ & $\left\{\sigma_1,-\sigma_1, \sigma_1,-\sigma_1\right\}$ & $\left\{-\sigma_2, \sigma_2,-\sigma_2, \sigma_2\right\}$ & $2$ \\
$G_1$ & $Y_1$ & $\left\{Y_{48}, Y_{24}\right\}$ & $\left\{Y_{15}, Y_{14}\right\}$ & $\left\{Y_{38}, Y_{39}\right\}$ & $\left\{-Y_{46},-Y_{47},-Y_{22},-Y_{23}\right\}$ & $\left\{-Y_{42},-Y_{43},-Y_{18},-Y_{19}\right\}$ & $Y_{25}$ \\
$G_2$ & $Y_1$ & $\left\{Y_{48}, Y_{24}\right\}$ & $\left\{-Y_{15},-Y_{14}\right\}$ & $\left\{-Y_{38},-Y_{39}\right\}$ & $\left\{Y_{46}, Y_{47}, Y_{22}, Y_{23}\right\}$ & $\left\{-Y_{42},-Y_{43},-Y_{18},-Y_{19}\right\}$ & $Y_{25}$ \\
\bottomrule
\end{tabular}
\label{tab:c4vd}
\end{table}

\begin{table}
\centering
\caption{Elements and irreps for group $C_{2v}$.}
\begin{tabular}{ccccc}
\toprule
$g$ & $E$ & $R_{16}$ & $I R_{22}$ & $I R_{17}$ \\
\midrule
$A_1$ & $1$ & $1$ & $1$ & $1$ \\
$A_2$ & $1$ & $1$ & $-1$ & $-1$ \\
$B_1$ & $1$ & $-1$ & $1$ & $-1$ \\
$B_2$ & $1$ & $-1$ & $-1$ & $1$ \\
\bottomrule
\end{tabular}
\label{tab:c2v}
\end{table}

\begin{table}
\centering
\caption{Elements and irreps for group $C_{2v}^D$.}
\begin{tabular}{cccccc}
\toprule
$g$ & $E$ & $\left\{R_{40}, R_{16}\right\}$ & $\left\{I R_{41}, I R_{17}\right\}$ & $\left\{I R_{46}, I R_{22}\right\}$ & $R_{25}$ \\
\midrule
$A_1$ & $1$ & $1$ & $1$ & $1$ & $1$ \\
$A_2$ & $1$ & $1$ & $-1$ & $-1$ & $1$ \\
$B_1$ & $1$ & $-1$ & $-1$ & $1$ & $1$ \\
$B_2$ & $1$ & $-1$ & $1$ & $-1$ & $1$ \\
$G$ & $2$ & $0$ & $0$ & $0$ & $-2$ \\
\midrule
$G$ & $Y_1$ & $\left\{Y_{40}, Y_{16}\right\}$ & $\left\{-Y_{41},-Y_{17}\right\}$ & $\left\{-Y_{46},-Y_{22}\right\}$ & $Y_{25}$ \\
\bottomrule
\end{tabular}
\label{tab:c2vd}
\end{table}

\begin{table}
\centering
\caption{Elements and irreps for group $C_{3v}$.}
\begin{tabular}{cccc}
\toprule
$g$ & $E$ & $\left\{R_2, R_3\right\}$ & $\left\{I R_{17}, I R_{19}, I R_{21}\right\}$ \\
$A_1$ & $1$ & $1$ & $1$ \\
$A_2$ & $1$ & $1$ & $-1$ \\
$E$ & $2$ & $-1$ & $0$ \\
\midrule
$E$ & $X_1$ & $\left\{X_{2}, X_{3}\right\}$ & $\left\{-X_{4},-X_{5},-X_{6}\right\}$ \\
\bottomrule
\end{tabular}
\label{tab:c3v}
\end{table}

\begin{table}
\centering
\caption{Elements and irreps for group $C_{3v}^D$.}
\begin{tabular}{ccccccc}
\toprule
$g$ & $E$ & $\left\{R_{3}, R_{2}\right\}$ & $\left\{R_{26}, R_{27}\right\}$ & $\left\{I R_{41}, I R_{19}, I R_{21}\right\}$ & $\left\{I R_{43}, I R_{45}, I R_{17}\right\}$ & $R_{25}$ \\
\midrule
$A_1$ & $1$ & $1$ & $1$ & $1$ & $1$ & $1$ \\
$A_2$ & $1$ & $1$ & $1$ & $-1$ & $-1$ & $1$ \\
$K_1$ & $1$ & $-1$ & $1$ & $i$ & $-i$ & $-1$ \\
$K_2$ & $1$ & $-1$ & $1$ & $-i$ & $i$ & $-1$ \\
$E$ & $2$ & $-1$ & $-1$ & $0$ & $0$ & $2$ \\
$G$ & $2$ & $1$ & $-1$ & $0$ & $0$ & $-2$ \\
\midrule
$E$ & $X_1$ & $\left\{X_3, X_2\right\}$ & $\left\{X_2, X_3\right\}$ & $\left\{-X_4,-X_5,-X_6\right\}$ & $\left\{-X_5,-X_6,-X_4\right\}$ & $X_1$ \\
$G$ & $Y_1$ & $\left\{Y_{3}, Y_{2}\right\}$ & $\left\{Y_{26}, Y_{27}\right\}$ & $\left\{-Y_{41},-Y_{19},-Y_{21}\right\}$ & $\left\{-Y_{43},-Y_{45},-Y_{17}\right\}$ & $Y_{25}$ \\
\bottomrule
\end{tabular}
\label{tab:c3vd}
\end{table}

\begin{table}
\centering
\caption{Elements and irreps for group $C_{2}$.}
\begin{tabular}{ccc}
\toprule
$g (nm0)$ & $E$ & $I R_{24}$ \\
$g (nnm)$ & $E$ & $I R_{17}$ \\
\midrule
$A$ & $1$ & $1$ \\
$B$ & $1$ & $-1$ \\
\bottomrule
\end{tabular}
\label{tab:c2}
\end{table}

\begin{table}
\centering
\caption{Elements and irreps for group $C_{2}^D$.}
\begin{tabular}{ccccc}
\toprule
$g (nm0)$ & $E$ & $I R_{48}$ & $R_{25}$ & $I R_{24}$ \\
$g (nnm)$ & $E$ & $I R_{41}$ & $R_{25}$ & $I R_{17}$ \\
\midrule
$A_1$ & $1$ & $1$ & $1$ & $1$ \\ 
$A_2$ & $1$ & $-1$ & $1$ & $-1$ \\
$K_1$ & $1$ & $i$ & $-1$ & $-i$ \\
$K_2$ & $1$ & $-i$ & $-1$ & $i$ \\
\bottomrule
\end{tabular}
\label{tab:c2d}
\end{table}

\begin{table}
\centering
\caption{Elements and irreps for group $C_{1}$.}
\begin{tabular}{cc}
\toprule
$g$ & $E$ \\
\midrule
$A$ & $1$ \\
\bottomrule
\end{tabular}
\label{tab:c1}
\end{table}

\begin{table}
\centering
\caption{Elements and irreps for group $C_{1}^D$.}
\begin{tabular}{ccc}
\toprule
$g$ & $E$ & $R_{25}$ \\
\midrule
$A$ & $1$ & $1$ \\ 
$K$ & $1$ & $-1$ \\
\bottomrule
\end{tabular}
\label{tab:c1d}
\end{table}

\section{The construction of the one-hadron operators}
\label{sec:one}
In this paper, we focus on the construction of meson-like operators. For a comprehensive treatment of single-baryon operator construction, we refer the reader to Refs.~\cite{Basak:2005ir, Basak:2005aq, Basak:2007kj}. Single-meson operators can be constructed using a quark bilinear of the form $\psi^{\prime} \Gamma \psi$, where the quantum numbers are determined by the choice of the gamma matrix $\Gamma$. For example, $\sum_{\vec{x}} e^{-i \vec{p} \cdot \vec{x}} \bar{d}(\vec{x}) \gamma_5 u(\vec{x})$ interpolate a $\pi^+$ with momentum $\vec{p}$, while $\sum_{f, \vec{x}} Q_f \bar{\psi}_f(\vec{x}) \gamma_t \psi_f(\vec{x})$ corresponds to the electromagnetic current and measures the electric charge of a hadron. To interpolate states with higher spins or quantum numbers that are inaccessible through simple gamma matrices, it is useful to introduce spatial separation between the quarks and connect them via gauge links. A convenient technique is to insert covariant derivatives between the quarks~\cite{Dudek:2010wm, Thomas:2011rh}, as employed by many studies, see e.g., Refs.~\cite{Wilson:2023hzu, Wilson:2023anv, Gayer:2021xzv, Yeo:2024chk, Whyte:2024ihh}. For alternative techniques, see Ref.~\cite{Morningstar:1999rf}.

In previous methods, operators with good quantum numbers $|jm\rangle$ were first projected onto the helicity basis, a process that involves the complexity of rotating operators into states with well-defined $|j\lambda\rangle$ on a finite volume. This requires a two-stage rotational procedure and the selection of a reference direction. Subsequently, the helicity operators were “subduced” into the irreps of the lattice group, which entails calculating Clebsch-Gordan coefficients for each lattice group individually. In this work, we adopt a more straightforward approach, directly projecting operators into the irreps of the lattice groups. This method can also be extended for use in $N$-hadron projections.

\subsection{Construction method}
In the $SU(2)$ group, if an operator has well-defined $|jm \rangle$, it will transform covariantly under the group element $g$ as
\begin{equation}
g A_{jm} g^{\dagger} = (-1)^{P} \sum_{m^{\prime}} A_{jm^{\prime}} \mathcal{D}_{m^{\prime}m}^{j}(g),
\label{eq:trans}
\end{equation}
where $J=0$ for scalar and pseudoscalar mesons, $J=1$ for vector and axial vector mesons, and $J=\frac{1}{2}$ for nucleons. $(-1)^{P}$ accounts for the internal $P$-parity of the particle. For vector-like mesons, suppose that the operator $A_x, A_y, A_z$ forms a Cartesian vector, then we can build the rank-$1$ irreducible tensor $A_{1m}$ by
\begin{equation}
A_{1, \pm1} = \mp \frac{i}{\sqrt{2}}(A_x \mp i A_y), \, A_{1,0} = i A_z
\label{eq:ir_tensor}
\end{equation}
such that $A_{1m}$ satisfies Eq.~\ref{eq:trans} with $j=1$. In the same way, one could write down irreducible tensors with higher ranks. The basic building blocks, the gamma matrices $\Gamma_{1m}$ and the covariant derivatives $\overleftrightarrow{{\mathcal{D}}}_{1m}$ are written in the irreducible tensor form as in Eq.~\ref{eq:ir_tensor}.

This process is equivalent to what was done in Ref.~\cite{Dudek:2010wm, Thomas:2011rh}, which is formally
\begin{equation}
O_{J M}(\vec{P}) = \sum_{i} \operatorname{CGs}(m_i) \times \sum_{\vec{x}} e^{-i \vec{P} \cdot \vec{x}} \bar{\psi}^{\prime}(\vec{x}, t) \Gamma_{m_1} \prod_{i>1} \overleftrightarrow{\mathcal{D}}_{m_i} \psi(\vec{x}, t).
\label{eq:OJM}
\end{equation}

Instead of constructing the helicity operators, we apply the irrep projection directly from the continuum operators, see e.g., Ref.~\cite{Prelovsek:2016iyo}. The desired operators with specific irrep $\Gamma$, row $\mu$, with CM momentum $\vec{P}$ are given by
\begin{equation}
O_{\Gamma, \mu}(\vec{P}) = \sum_{g \in G[\vec{P}]} T_{\mu, \mu}^{\Gamma}(g) \, g O(\vec{P}) g^{\dagger},
\label{eq:projection}
\end{equation}
where $T_{\mu, \mu}^{\Gamma}$ is the diagonal matrix elements of the irrep $\Gamma$, and $g$ loops over all elements in the (little) group $G$. The relation between $G$ and the CM momenta is tabulated in Tab.~\ref{tab:group}. $O(\vec{P})$ is an operator with arbitrary form, and is here chosen to be $O_{J M}(\vec{p})$ in Eq.~\ref{eq:OJM}. We can obtain all operators under a given maximum number of covariant derivatives by scanning the number of covariant derivatives $0 \leq n_D\ \leq N_{\mathcal{D}}$, the type of gamma matrices, the total spin in the rest frame at the infinite volume $|J^{\Gamma}-n_{\mathcal{D}}| \leq J \leq J^{\Gamma}+n_{\mathcal{D}}$. $J^{\Gamma}$ denotes the spin coming from the gamma matrix.

\subsection{List of one-meson operators}
The one-hadron operators constructed using the gamma matrices $S$, $P$, $V$, and $A$ correspond to the insertion of scalar, pseudoscalar, vector, and axial-vector gamma matrices, respectively. In this work, all operators are projected to the first row of the irrep for convenience. Also, we present only operators containing zero or one covariant derivative. The operators constructed for CM momenta of the types in $\Omega$ are tabulated in Tab.~\ref{tab:one-000}, \ref{tab:one-00n}, \ref{tab:one-0nn}, \ref{tab:one-nnn}, \ref{tab:one-nm0}, \ref{tab:one-nnm}, and \ref{tab:one-nmp}, respectively.

In the following, we assume that the gamma matrices and covariant derivatives are sandwiched between $\bar{\psi}^{\prime}$ and $\psi$, with arrows on the derivatives omitted for simplicity. The listed operators are presented such that the overall constants appear in front of the operators is $1$. For example, an operator denoted as $P\mathcal{D}_x$ from the $T_1^+$ irrep of the $O_h$ group refers to $\bar{\psi}^{\prime} \gamma_5 \overleftrightarrow{\mathcal{D}}_x \psi$ or $\bar{\psi}^{\prime} \gamma_4\gamma_5 \overleftrightarrow{\mathcal{D}}_x \psi$.

\begin{table}[htbp]
\centering
\caption{The list of one-hadron operators projected onto the irreps of the cubic group $O_h$. The gamma matrices $S$, $P$, $V$, and $A$ correspond to the scalar, pseudoscalar, vector, and axial-vector types, respectively. Arrows on the derivatives have been omitted for simplicity, and overall constants have been ignored.}
\addtolength{\tabcolsep}{6pt}
\begin{tabular}{cc}
\toprule
irrep & operator \\
\midrule
\multirow{2}{*}{$A_1^+$} & $S$ \\
& $V_x\mathcal{D}_x + V_y\mathcal{D}_y + V_z\mathcal{D}_z$ \\
\midrule
\multirow{2}{*}{$A_1^-$} & $P$ \\
& $A_x\mathcal{D}_x + A_y\mathcal{D}_y + A_z\mathcal{D}_z$ \\
\midrule
\multirow{1}{*}{$A_2^+$} & - \\
\midrule
\multirow{1}{*}{$A_2^-$} & - \\
\midrule
\multirow{1}{*}{$E^+$} & $V_x\mathcal{D}_x + V_y\mathcal{D}_y -2 V_z\mathcal{D}_z$ \\
\midrule
\multirow{1}{*}{$E^-$} & $A_x\mathcal{D}_x + A_y\mathcal{D}_y -2 A_z\mathcal{D}_z$ \\
\midrule
\multirow{3}{*}{$T_1^+$} & $A_x$ \\
& $P\mathcal{D}_x$ \\
& $-V_z\mathcal{D}_y + V_y\mathcal{D}_z$ \\
\midrule
\multirow{3}{*}{$T_1^-$} & $V_x$ \\
& $SD_x$ \\
& $A_z\mathcal{D}_y - A_y\mathcal{D}_z$ \\
\midrule
\multirow{1}{*}{$T_2^+$} & $V_z\mathcal{D}_y + V_y\mathcal{D}_z$ \\
\midrule
\multirow{1}{*}{$T_2^-$} & $A_z\mathcal{D}_y + A_y\mathcal{D}_z$ \\
\bottomrule
\end{tabular}
\addtolength{\tabcolsep}{-6pt}
\label{tab:one-000}
\end{table}

\begin{table}[htbp]
\centering
\caption{The one-hadron operator list for group $C_{4v}$ with the CM momentum $[0,0,n]$. Notations as in Tab.~\ref{tab:one-000}.}
\addtolength{\tabcolsep}{6pt}
\begin{tabular}{cc}
\toprule
irrep & operator \\
\midrule
\multirow{6}{*}{$A_1$} & $S$ \\
& $V_z$ \\
& $S\mathcal{D}_z$ \\
& $V_x\mathcal{D}_x + V_y\mathcal{D}_y + V_z\mathcal{D}_z$ \\
& $V_x\mathcal{D}_x + V_y\mathcal{D}_y -2 V_z\mathcal{D}_z$ \\
& $A_y\mathcal{D}_x - A_x\mathcal{D}_y$ \\
\midrule
\multirow{6}{*}{$A_2$} & $P$ \\
& $A_z$ \\
& $P\mathcal{D}_z$ \\
& $A_x\mathcal{D}_x + A_y\mathcal{D}_y + A_z\mathcal{D}_z$ \\
& $A_x\mathcal{D}_x + A_y\mathcal{D}_y -2 A_z\mathcal{D}_z$ \\
& $V_y\mathcal{D}_x - V_x\mathcal{D}_y$ \\
\midrule
\multirow{2}{*}{$B_1$} & $V_x\mathcal{D}_x - V_y\mathcal{D}_y$ \\
& $A_y\mathcal{D}_x + A_x\mathcal{D}_y$ \\
\midrule
\multirow{2}{*}{$B_2$} & $A_x\mathcal{D}_x - A_y\mathcal{D}_y$ \\
& $V_y\mathcal{D}_x + V_x\mathcal{D}_y$ \\
\midrule
\multirow{8}{*}{$E$} & $V_y$ \\
& $A_x$ \\
& $S\mathcal{D}_y$ \\
& $P\mathcal{D}_x$ \\
& $-V_z\mathcal{D}_y + V_y\mathcal{D}_z$ \\
& $V_z\mathcal{D}_y + V_y\mathcal{D}_z$ \\
& $-A_z\mathcal{D}_x + A_x\mathcal{D}_z$ \\
& $A_z\mathcal{D}_x + A_x\mathcal{D}_z$ \\
\bottomrule
\end{tabular}
\addtolength{\tabcolsep}{-6pt}
\label{tab:one-00n}
\end{table}

\begin{table}[htbp]
\centering
\caption{The one-hadron operator list for group $C_{2v}$ with the CM momentum $[0,n,n]$. Notations as in Tab.~\ref{tab:one-000}.}
\addtolength{\tabcolsep}{6pt}
\begin{tabular}{cc}
\toprule
irrep & operator \\
\midrule
\multirow{8}{*}{$A_1$} & $S$ \\
& $V_y+V_z$ \\
& $S (\mathcal{D}_y + \mathcal{D}_z)$ \\
& $V_x\mathcal{D}_x + V_y\mathcal{D}_y + V_z\mathcal{D}_z$ \\
& $2 V_x\mathcal{D}_x - V_y\mathcal{D}_y - V_z\mathcal{D}_z$ \\
& $V_z\mathcal{D}_y + V_y\mathcal{D}_z$ \\
& $A_y\mathcal{D}_x - A_z\mathcal{D}_x - A_x\mathcal{D}_y + A_x\mathcal{D}_z$ \\
& $A_y\mathcal{D}_x - A_z\mathcal{D}_x + A_x\mathcal{D}_y - A_x\mathcal{D}_z$ \\
\midrule
\multirow{8}{*}{$A_2$} & $P$ \\
& $A_y+A_z$ \\
& $P (\mathcal{D}_y + \mathcal{D}_z)$ \\
& $A_x\mathcal{D}_x + A_y\mathcal{D}_y + A_z\mathcal{D}_z$ \\
& $2 A_x\mathcal{D}_x - A_y\mathcal{D}_y - A_z\mathcal{D}_z$ \\
& $A_z\mathcal{D}_y + A_y\mathcal{D}_z$ \\
& $V_y\mathcal{D}_x - V_z\mathcal{D}_x - V_x\mathcal{D}_y + V_x\mathcal{D}_z$ \\
& $V_y\mathcal{D}_x - V_z\mathcal{D}_x + V_x\mathcal{D}_y - V_x\mathcal{D}_z$ \\
\midrule
\multirow{8}{*}{$B_1$} & $V_y - V_z$ \\
& $A_x$ \\
& $P\mathcal{D}_x$ \\
& $S (\mathcal{D}_y - \mathcal{D}_z)$ \\
& $V_z\mathcal{D}_y - V_y\mathcal{D}_z$ \\
& $V_y\mathcal{D}_y - V_z\mathcal{D}_z$ \\
& $A_y\mathcal{D}_x + A_z\mathcal{D}_x - A_x\mathcal{D}_y - A_x\mathcal{D}_z$ \\
& $A_y\mathcal{D}_x + A_z\mathcal{D}_x + A_x\mathcal{D}_y + A_x\mathcal{D}_z$ \\
\midrule
\multirow{8}{*}{$B_2$} & $A_y - A_z$ \\
& $V_x$ \\
& $S\mathcal{D}_x$ \\
& $P (\mathcal{D}_y - \mathcal{D}_z)$ \\
& $A_z\mathcal{D}_y - A_y\mathcal{D}_z$ \\
& $A_y\mathcal{D}_y - A_z\mathcal{D}_z$ \\
& $V_y\mathcal{D}_x + V_z\mathcal{D}_x - V_x\mathcal{D}_y - V_x\mathcal{D}_z$ \\
& $V_y\mathcal{D}_x + V_z\mathcal{D}_x + V_x\mathcal{D}_y + V_x\mathcal{D}_z$ \\
\bottomrule
\end{tabular}
\addtolength{\tabcolsep}{-6pt}
\label{tab:one-0nn}
\end{table}

\begin{table}[htbp]
\centering
\caption{The one-hadron operator list for group $C_{3v}$ with the CM momentum $[n,n,n]$. Notations as in Tab.~\ref{tab:one-000}.}
\addtolength{\tabcolsep}{6pt}
\begin{tabular}{cc}
\toprule
irrep & operator \\
\midrule
\multirow{6}{*}{$A_1$} & $S$ \\
& $V_x + V_y + V_z$ \\
& $S (\mathcal{D}_x + \mathcal{D}_y + \mathcal{D}_z)$ \\
& $V_x\mathcal{D}_x + V_y\mathcal{D}_y + V_z\mathcal{D}_z$ \\
& $V_y\mathcal{D}_x + V_z\mathcal{D}_x + V_x\mathcal{D}_y + V_z\mathcal{D}_y + V_x\mathcal{D}_z + V_y\mathcal{D}_z$ \\
& $A_y\mathcal{D}_x - A_z\mathcal{D}_x - A_x\mathcal{D}_y + A_z\mathcal{D}_y + A_x\mathcal{D}_z - A_y\mathcal{D}_z$ \\
\midrule
\multirow{6}{*}{$A_2$} & $P$ \\
& $A_x + A_y + A_z$ \\
& $P (\mathcal{D}_x + \mathcal{D}_y + \mathcal{D}_z)$ \\
& $A_x\mathcal{D}_x + A_y\mathcal{D}_y + A_z\mathcal{D}_z$ \\
& $A_y\mathcal{D}_x + A_z\mathcal{D}_x + A_x\mathcal{D}_y + A_z\mathcal{D}_y + A_x\mathcal{D}_z + A_y\mathcal{D}_z$ \\
& $V_y\mathcal{D}_x - V_z\mathcal{D}_x - V_x\mathcal{D}_y + V_z\mathcal{D}_y + V_x\mathcal{D}_z - V_y\mathcal{D}_z$ \\
\midrule
\multirow{10}{*}{$E$} & $V_x - V_y$ \\
& $A_x + A_y -2 A_z$ \\
& $S (\mathcal{D}_x - \mathcal{D}_y)$ \\
& $P (\mathcal{D}_x + \mathcal{D}_y - 2 \mathcal{D}_z)$ \\
& $-2 V_y\mathcal{D}_x - V_z\mathcal{D}_x +2 V_x\mathcal{D}_y + V_z\mathcal{D}_y + V_x\mathcal{D}_z - V_y\mathcal{D}_z$ \\
& $V_x\mathcal{D}_x - V_y\mathcal{D}_y$ \\
& $V_z\mathcal{D}_x - V_z\mathcal{D}_y + V_x\mathcal{D}_z - V_y\mathcal{D}_z$ \\
& $A_z\mathcal{D}_x + A_z\mathcal{D}_y - A_x\mathcal{D}_z - A_y\mathcal{D}_z$ \\
& $2 A_y\mathcal{D}_x - A_z\mathcal{D}_x +2 A_x\mathcal{D}_y - A_z\mathcal{D}_y - A_x\mathcal{D}_z - A_y\mathcal{D}_z$ \\
& $A_x\mathcal{D}_x + A_y\mathcal{D}_y -2 A_z\mathcal{D}_z$ \\
\bottomrule
\end{tabular}
\addtolength{\tabcolsep}{-6pt}
\label{tab:one-nnn}
\end{table}

\begin{table}[htbp]
\centering
\caption{The one-hadron operator list for group $C_{2}$ with the CM momentum $[n,m,0]$. Notations as in Tab.~\ref{tab:one-000}.}
\addtolength{\tabcolsep}{6pt}
\begin{tabular}{cc}
\toprule
irrep & operator \\
\midrule
\multirow{16}{*}{$A$} & $S$ \\
& $V_x$ \\
& $V_y$ \\
& $A_z$ \\
& $S\mathcal{D}_x$ \\
& $S\mathcal{D}_y$ \\
& $P\mathcal{D}_z$ \\
& $V_x\mathcal{D}_x + V_y\mathcal{D}_y + V_z\mathcal{D}_z$ \\
& $V_y\mathcal{D}_x$ \\
& $V_x\mathcal{D}_x - V_y\mathcal{D}_y$ \\
& $V_x\mathcal{D}_y$ \\
& $V_z\mathcal{D}_z$ \\
& $A_x\mathcal{D}_z$ \\
& $A_z\mathcal{D}_x$ \\
& $A_y\mathcal{D}_z$ \\
& $A_z\mathcal{D}_y$ \\
\midrule
\multirow{16}{*}{$B$} & $P$ \\
& $A_x$ \\
& $A_y$ \\
& $V_z$ \\
& $P\mathcal{D}_x$ \\
& $P\mathcal{D}_y$ \\
& $S\mathcal{D}_z$ \\
& $A_x\mathcal{D}_x + A_y\mathcal{D}_y + A_z\mathcal{D}_z$ \\
& $A_y\mathcal{D}_x$ \\
& $A_x\mathcal{D}_x - A_y\mathcal{D}_y$ \\
& $A_x\mathcal{D}_y$ \\
& $A_z\mathcal{D}_z$ \\
& $V_x\mathcal{D}_z$ \\
& $V_z\mathcal{D}_x$ \\
& $V_y\mathcal{D}_z$ \\
& $V_z\mathcal{D}_y$ \\
\bottomrule
\end{tabular}
\addtolength{\tabcolsep}{-6pt}
\label{tab:one-nm0}
\end{table}

\begin{table}[htbp]
\centering
\caption{The one-hadron operator list for group $C_{2}$ with the CM momentum $[n,n,m]$. Notations as in Tab.~\ref{tab:one-000}.}
\addtolength{\tabcolsep}{6pt}
\begin{tabular}{cc}
\toprule
irrep & operator \\
\midrule
\multirow{16}{*}{$A$} & $S$ \\
& $V_x + V_y$ \\
& $V_z$ \\
& $A_x - A_y$ \\
& $S(D_x + D_y)$ \\
& $SD_z$ \\
& $P(D_x - D_y)$ \\
& $V_x\mathcal{D}_x + V_y\mathcal{D}_y$ \\
& $V_z\mathcal{D}_z$ \\
& $V_x\mathcal{D}_z + V_y\mathcal{D}_z$ \\
& $V_z\mathcal{D}_x + V_z\mathcal{D}_y$ \\
& $V_y\mathcal{D}_x + V_x\mathcal{D}_y$ \\
& $A_z\mathcal{D}_x - A_z\mathcal{D}_y$ \\
& $A_x\mathcal{D}_z - A_y\mathcal{D}_z$ \\
& $A_y\mathcal{D}_x - A_x\mathcal{D}_y$ \\
& $A_x\mathcal{D}_x - A_y\mathcal{D}_y$ \\
\midrule
\multirow{16}{*}{$B$} & $P$ \\
& $A_x + A_y$ \\
& $A_z$ \\
& $V_x - V_y$ \\
& $P(D_x + D_y)$ \\
& $PD_z$ \\
& $S(D_x - D_y)$ \\
& $A_x\mathcal{D}_x + A_y\mathcal{D}_y$ \\
& $A_z\mathcal{D}_z$ \\
& $A_x\mathcal{D}_z + A_y\mathcal{D}_z$ \\
& $A_z\mathcal{D}_x + A_z\mathcal{D}_y$ \\
& $A_y\mathcal{D}_x + A_x\mathcal{D}_y$ \\
& $V_z\mathcal{D}_x - V_z\mathcal{D}_y$ \\
& $V_x\mathcal{D}_z - V_y\mathcal{D}_z$ \\
& $V_y\mathcal{D}_x - V_x\mathcal{D}_y$ \\
& $V_x\mathcal{D}_x - V_y\mathcal{D}_y$ \\
\bottomrule
\end{tabular}
\addtolength{\tabcolsep}{-6pt}
\label{tab:one-nnm}
\end{table}

\begin{table}[htbp]
\centering
\caption{The one-hadron operator list for group $C_{1}$ with the CM momentum $[n,m,p]$. Notations as in Tab.~\ref{tab:one-000}.}
\addtolength{\tabcolsep}{6pt}
\begin{tabular}{cc}
\toprule
irrep & operator \\
\midrule
\multirow{4}{*}{$A$} & $S, P, V_x, V_y, V_z, A_x, A_y, A_z$ \\
& $S\mathcal{D}_x, S\mathcal{D}_y, S\mathcal{D}_z, P\mathcal{D}_x, P\mathcal{D}_y, P\mathcal{D}_z$ \\
& $V_x\mathcal{D}_x, V_y\mathcal{D}_y, V_z\mathcal{D}_z, V_x\mathcal{D}_y, V_y\mathcal{D}_x, V_x\mathcal{D}_z, V_z\mathcal{D}_x, V_y\mathcal{D}_z, V_z\mathcal{D}_y$ \\
& $A_x\mathcal{D}_x, A_y\mathcal{D}_y, A_z\mathcal{D}_z, A_x\mathcal{D}_y, A_y\mathcal{D}_x, A_x\mathcal{D}_z, A_z\mathcal{D}_x, A_y\mathcal{D}_z, A_z\mathcal{D}_y$ \\
\bottomrule
\end{tabular}
\addtolength{\tabcolsep}{-6pt}
\label{tab:one-nmp}
\end{table}

Note that there are no operators with $N_{\mathcal{D}} \leq 1$ for the irrep $A_2^{\pm}$ of the $O_h$ group. For the $C_2$ and $C_1$ groups, we have applied linear combinations of results derived from Eq.~\ref{eq:projection} to simplify the expressions. As the symmetry decreases from $O_h$ to the trivial group $C_1 = \{E\}$, the number of irreps (or classes) reduces, leading to greater mixing among operators. Specifically, for momentum $\vec{P} = [0,0,0]$, spatial inversion is a symmetry operation of the $O_h$ group, making $P$-parity a conserved quantum number. Consequently, operators are classified by parity, and different parities do not mix. For example, the scalar operator $S \sim \bar{\psi}^{\prime} \gamma_5 \psi$ in $A_1^+$ does not mix with $V_x \sim \bar{\psi}^{\prime} \gamma_x \psi$ in $T_1^-$. However, for momentum $\vec{P} = [0,0,n]$, spatial inversion is no longer a symmetry operation due to the definition of an axis, allowing both $S$ and $V_z$ to interpolate states in the $A_1$ irrep. Furthermore, for $\vec{P} = [n,m,p]$, where no spatial symmetry remains, all operators of any type can mix. In this case, all states in the infinite volume with different quantum numbers, including bound states, virtual states, resonances, and scattering states, appear in the single irrep $A$.

As a consistency check, we verify the number of operators in each group. The number of quark bilinear operators built purely from gamma matrices is $1, 1, 3, 3$ for the $S, P, V, A$ types, where the axial vector ($A$) and vector ($V$) operators can each take three spatial directions. Introducing one covariant derivative, the number of operators becomes $(1 + 1 + 3 + 3) \times 3$ for the three possible directions of the covariant derivative. Thus, we expect a total of $(1 + 1 + 3 + 3) \times 4 = 32$ operators when considering $N_{\mathcal{D}} \leq 1$.

Since we do not explicitly enumerate the operators for irreps with $\mu \neq 1$, we must multiply the number of operators for each irrep by the dimension of that irrep to obtain the full operator count. For instance, in the case of the group $C_{4v}$, the irreps $A_1, A_2, B_1,$ and $B_2$  are all one-dimensional, while the irrep $E$ is two-dimensional. the total number of operators is therefore given by $6 + 6 + 2 + 2 + 8 \times 2 = 32$, which yields the expected count. This confirms that the projection method employed exhausts the set of operators allowed by the symmetry group.

\section{The construction of the $N$-hadron operators}
\label{sec:two}
\subsection{Construction method}
To construct multi-body operators, we let $O(\vec{P})$ in Eq.~\ref{eq:projection} designate the $N$-body operators that are built from one-hadron operators in the infinite volume,
\begin{equation}
    O(\vec{P}) = \prod_{i=1}^{N} O_i(\vec{k}_i),
\end{equation}
where $O_i$ is the meson or baryon operator. The projection method is then applied to an arbitrary number of particles and non-zero CM momentum:
\begin{equation}
O_{\Gamma, \mu}(\vec{P}) = \sum_{g \in G[\vec{P}]} T_{\mu, \mu}^{\Gamma}(g) \, \prod_{i=1}^{N} (g O_i(\vec{k}_i) g^{\dagger}),
\end{equation}
and the transformation property $g O_i(\vec{k}_i) g^{\dagger}$ is read from Eq.~\ref{eq:trans}. Since it is not necessary to project the one-hadron operator to the irrep of the lattice group for constructing multi-particle states, the building blocks $O_i(\vec{k}_i)$ take the form of a one-hadron operator $O_{J M}$ from the continuum symmetry as in Eq.~\ref{eq:OJM}.

The total momentum has to be conserved $\sum_i \vec{k}_i = \vec{P}$, where $\vec{k}_i$ is the momentum for each individual hadron.

Note that the projected operator $O_{\Gamma, \mu}(\vec{P})$ will have the proper momentum configurations in the irrep, and the specific direction of $\vec{k}_i$ does not matter and will be one of the terms in the resulting operator. In principle, there are no restrictions on the angular momentum or helicity quantum number of $O(\vec{P})$ as long as it has a non-zero component in the $\Gamma$ irrep.

To build a space with a complete set of operators with the given quantum number under certain $\max{|\vec{k}_i|}$, we systematically scan all $\vec{k}_i$ vectors within the given momentum shell. To avoid generating redundant or linearly dependent operators, we build an $(N_{\text{op}} + 1) \times N_{\text{mon}}$ matrix, where $N_{\text{op}}$ is the number of the operator already built, and $N_{\text{mon}}$ the number of different $O_i(\vec{k}_i)$ monomial. Let $M_{ij}$ denote the coefficient in front of the $j$th $O_i(\vec{k}_i)$ monomial for the operator $O_{\Gamma, \mu}^{(i)}(\vec{P})$. The first $N_{\text{op}}$-th rows are for the operator list that are known to be linearly independent, and the $(N_{\text{op}} + 1)$-th row is the pending constructed operator. Let $M_{ij}^{\prime}$ denote the submatrix containing only the first $N{\text{op}}$ rows. The new operator is accepted into the operator basis if and only if $\operatorname{Rank}(M_{ij}) > \operatorname{Rank}(M_{ij}^{\prime})$,
which indicates linear independence from the existing set.

It is often advantageous to apply linear transformations to the operator basis for physical interpretation. For example, in the rest frame $\vec{P} = 0$, the operators can be combined according to definite partial waves and couple predominantly to the low partial waves of interest. This partial-wave projection technique, as discussed in Ref.~\cite{Prelovsek:2016iyo}, is also implemented in our \texttt{Mathematica} package~\cite{github}.

\subsection{List of two-hadron operators}
\label{sec:list_two}
In Tab.~\ref{tab:mm-000-A1+}, we present the complete set of meson-meson operators live in the $A_1^+$ irrep of $O_h$ group in the rest frame, constructed up to one unit of relative momentum (in units of $2 \pi / L$). The table assumes the particles are distinguishable. These operators can be directly applied to the study of the $S$ wave $D\pi$ scattering and the $D_0^*(2300)$ resonance~\cite{Yan:2023gvq, Yan:2024yuq}, as well as $\pi\pi$, $K\pi$, scattering, etc. Operators in the rest frame provide the most direct and valuable input for determining the scattering amplitude.

Each symbol $P, S, V, A$ denotes a one-hadron operator constructed as described in Section~\ref{sec:one}, potentially including any number of covariant derivatives, provided it carries definite angular momentum quantum numbers $J, M$. Additional operator lists for other little groups and irreps—including meson-meson, meson-baryon, and baryon-baryon combinations—are compiled in Appendix~\ref{sec:append_list_mm}, \ref{sec:append_list_bb}, and \ref{sec:append_list_mb}, intended to serve as a practical reference for future spectroscopy and matrix element studies.

\begin{table}[htbp]
\centering
\caption{The list of two-hadron operators projected onto the $A_1^+$ irrep of the cubic group $O_h$. The gamma matrices $S$, $P$, $V$, and $A$ correspond to the scalar, pseudoscalar, vector, and axial-vector types, respectively. Arrows on the derivatives have been omitted for simplicity, and overall constants have been ignored.}
\addtolength{\tabcolsep}{6pt}
\begin{tabular}{c}
\toprule
$P_1(0) P_2(0)$ \\
\midrule
$P_1(e_{x}) P_2(-e_{x}) + P_1(-e_{x}) P_2(e_{x}) + P_1(e_{y}) P_2(-e_{y}) + P_1(-e_{y}) P_2(e_{y})$ \\
$+ P_1(e_{z}) P_2(-e_{z}) + P_1(-e_{z}) P_2(e_{z})$ \\
\midrule
$P_2(e_{x}) A_{1,x}(-e_{x}) - P_2(-e_{x}) A_{1,x}(e_{x}) + P_2(e_{y}) A_{1,y}(-e_{y}) - P_2(-e_{y}) A_{1,y}(e_{y})$ \\
$+ P_2(e_{z}) A_{1,z}(-e_{z}) - P_2(-e_{z}) A_{1,z}(e_{z})$ \\
\midrule
$S_1(0) S_2(0)$ \\
\midrule
$S_1(e_{x}) S_2(-e_{x}) + S_1(-e_{x}) S_2(e_{x}) + S_1(e_{y}) S_2(-e_{y}) + S_1(-e_{y}) S_2(e_{y})$ \\
$+ S_1(e_{z}) S_2(-e_{z}) + S_1(-e_{z}) S_2(e_{z})$ \\
\midrule
$S_1(e_{x}) V_{2,x}(-e_{x}) - S_1(-e_{x}) V_{2,x}(e_{x}) + S_1(e_{y}) V_{2,y}(-e_{y}) - S_1(-e_{y}) V_{2,y}(e_{y})$ \\
$+ S_1(e_{z}) V_{2,z}(-e_{z}) - S_1(-e_{z}) V_{2,z}(e_{z})$ \\
\midrule
$V_{1,x}(0) V_{2,x}(0) + V_{1,y}(0) V_{2,y}(0) + V_{1,z}(0) V_{2,z}(0)$ \\
\midrule
$2 V_{1,x}(e_{x}) V_{2,x}(-e_{x}) + 2 V_{1,x}(-e_{x}) V_{2,x}(e_{x}) - V_{1,x}(e_{y}) V_{2,x}(-e_{y}) - V_{1,x}(-e_{y}) V_{2,x}(e_{y})$ \\
$- V_{1,x}(e_{z}) V_{2,x}(-e_{z}) - V_{1,x}(-e_{z}) V_{2,x}(e_{z}) - V_{1,y}(e_{x}) V_{2,y}(-e_{x}) - V_{1,y}(-e_{x}) V_{2,y}(e_{x})$ \\
$+ 2 V_{1,y}(e_{y}) V_{2,y}(-e_{y}) + 2 V_{1,y}(-e_{y}) V_{2,y}(e_{y}) - V_{1,y}(e_{z}) V_{2,y}(-e_{z}) - V_{1,y}(-e_{z}) V_{2,y}(e_{z})$ \\
$- V_{1,z}(e_{x}) V_{2,z}(-e_{x}) - V_{1,z}(-e_{x}) V_{2,z}(e_{x}) - V_{1,z}(e_{y}) V_{2,z}(-e_{y}) - V_{1,z}(-e_{y}) V_{2,z}(e_{y})$ \\
$+ 2 V_{1,z}(e_{z}) V_{2,z}(-e_{z}) + 2 V_{1,z}(-e_{z}) V_{2,z}(e_{z})$ \\
\midrule
$2 V_{1,x}(e_{x}) V_{2,x}(-e_{x}) + 2 V_{1,x}(-e_{x}) V_{2,x}(e_{x}) + V_{1,x}(e_{y}) V_{2,x}(-e_{y}) + V_{1,x}(-e_{y}) V_{2,x}(e_{y})$ \\
$+ V_{1,x}(e_{z}) V_{2,x}(-e_{z}) + V_{1,x}(-e_{z}) V_{2,x}(e_{z}) + V_{1,y}(e_{x}) V_{2,y}(-e_{x}) + V_{1,y}(-e_{x}) V_{2,y}(e_{x})$ \\
$+ 2 V_{1,y}(e_{y}) V_{2,y}(-e_{y}) + 2 V_{1,y}(-e_{y}) V_{2,y}(e_{y}) + V_{1,y}(e_{z}) V_{2,y}(-e_{z}) + V_{1,y}(-e_{z}) V_{2,y}(e_{z})$ \\
$+ V_{1,z}(e_{x}) V_{2,z}(-e_{x}) + V_{1,z}(-e_{x}) V_{2,z}(e_{x}) + V_{1,z}(e_{y}) V_{2,z}(-e_{y}) + V_{1,z}(-e_{y}) V_{2,z}(e_{y})$ \\
$+ 2 V_{1,z}(e_{z}) V_{2,z}(-e_{z}) + 2 V_{1,z}(-e_{z}) V_{2,z}(e_{z})$ \\
\midrule
$A_{1,z}(e_{y}) V_{2,x}(-e_{y}) - A_{1,z}(-e_{y}) V_{2,x}(e_{y}) - A_{1,y}(e_{z}) V_{2,x}(-e_{z}) + A_{1,y}(-e_{z}) V_{2,x}(e_{z})$ \\
$- A_{1,z}(e_{x}) V_{2,y}(-e_{x}) + A_{1,z}(-e_{x}) V_{2,y}(e_{x}) + A_{1,x}(e_{z}) V_{2,y}(-e_{z}) - A_{1,x}(-e_{z}) V_{2,y}(e_{z})$ \\
$+ A_{1,y}(e_{x}) V_{2,z}(-e_{x}) - A_{1,y}(-e_{x}) V_{2,z}(e_{x}) - A_{1,x}(e_{y}) V_{2,z}(-e_{y}) + A_{1,x}(-e_{y}) V_{2,z}(e_{y})$ \\
\midrule
$A_{1,x}(0) A_{2,x}(0) + A_{1,y}(0) A_{2,y}(0) + A_{1,z}(0) A_{2,z}(0)$ \\
\midrule
$2 A_{1,x}(e_{x}) A_{2,x}(-e_{x}) + 2 A_{1,x}(-e_{x}) A_{2,x}(e_{x}) - A_{1,x}(e_{y}) A_{2,x}(-e_{y}) - A_{1,x}(-e_{y}) A_{2,x}(e_{y})$ \\
$- A_{1,x}(e_{z}) A_{2,x}(-e_{z}) - A_{1,x}(-e_{z}) A_{2,x}(e_{z}) - A_{1,y}(e_{x}) A_{2,y}(-e_{x}) - A_{1,y}(-e_{x}) A_{2,y}(e_{x})$ \\
$+ 2 A_{1,y}(e_{y}) A_{2,y}(-e_{y}) + 2 A_{1,y}(-e_{y}) A_{2,y}(e_{y}) - A_{1,y}(e_{z}) A_{2,y}(-e_{z}) - A_{1,y}(-e_{z}) A_{2,y}(e_{z})$ \\
$- A_{1,z}(e_{x}) A_{2,z}(-e_{x}) - A_{1,z}(-e_{x}) A_{2,z}(e_{x}) - A_{1,z}(e_{y}) A_{2,z}(-e_{y}) - A_{1,z}(-e_{y}) A_{2,z}(e_{y})$ \\
$+ 2 A_{1,z}(e_{z}) A_{2,z}(-e_{z}) + 2 A_{1,z}(-e_{z}) A_{2,z}(e_{z})$ \\
\midrule
$2 A_{1,x}(e_{x}) A_{2,x}(-e_{x}) + 2 A_{1,x}(-e_{x}) A_{2,x}(e_{x}) + A_{1,x}(e_{y}) A_{2,x}(-e_{y}) + A_{1,x}(-e_{y}) A_{2,x}(e_{y})$ \\
$+ A_{1,x}(e_{z}) A_{2,x}(-e_{z}) + A_{1,x}(-e_{z}) A_{2,x}(e_{z}) + A_{1,y}(e_{x}) A_{2,y}(-e_{x}) + A_{1,y}(-e_{x}) A_{2,y}(e_{x})$ \\
$+ 2 A_{1,y}(e_{y}) A_{2,y}(-e_{y}) + 2 A_{1,y}(-e_{y}) A_{2,y}(e_{y}) + A_{1,y}(e_{z}) A_{2,y}(-e_{z}) + A_{1,y}(-e_{z}) A_{2,y}(e_{z})$ \\
$+ A_{1,z}(e_{x}) A_{2,z}(-e_{x}) + A_{1,z}(-e_{x}) A_{2,z}(e_{x}) + A_{1,z}(e_{y}) A_{2,z}(-e_{y}) + A_{1,z}(-e_{y}) A_{2,z}(e_{y})$ \\
$+ 2 A_{1,z}(e_{z}) A_{2,z}(-e_{z}) + 2 A_{1,z}(-e_{z}) A_{2,z}(e_{z})$ \\
\bottomrule
\end{tabular}
\addtolength{\tabcolsep}{-6pt}
\label{tab:mm-000-A1+}
\end{table}

\subsection{Examples of multi-hadron operators}
Many questions in hadronic physics require a rigorous understanding of multi-particle systems. Among the most studied is the three-pion ($\pi\pi\pi$) system, which has been investigated in a variety of isospin channels, including maximal isospin~\cite{Hansen:2020otl, Mai:2018djl, Blanton:2019vdk, Culver:2019vvu, Fischer:2020jzp, Beane:2007es, Horz:2019rrn, Mai:2019fba, Brett:2021wyd}, isovector~\cite{Mai:2021nul}, and isoscalar~\cite{Yan:2024gwp} channels. Other three-meson systems, such as $KK\pi$, $K\pi\pi$~\cite{Dawid:2025zxc, Blanton:2021llb, Draper:2023boj}, and $KKK$~\cite{Alexandru:2020xqf}, have also attracted increasing attention. Additional insights have come from studies of systems with large isospin density~\cite{Abbott:2023coj, Detmold:2008fn, Detmold:2008yn, Detmold:2011kw} and from three-hadron systems involving charm quarks~\cite{Fu:2025joa}.

A particularly noteworthy example is the doubly charmed tetraquark candidate $T_{cc}$, which has been studied in several pioneering lattice works~\cite{Collins:2024sfi, Chen:2022vpo, Lyu:2023xro, Padmanath:2022cvl, Prelovsek:2025vbr, Du:2023hlu, Meng:2023bmz} via two-body $DD^*$ scattering analysis. However, a full understanding of this state requires addressing complications from nearby left-hand cuts~\cite{Prelovsek:2025vbr, Du:2023hlu, Meng:2023bmz} and the presence of the three-body $DD\pi$ channel~\cite{Dawid:2024dgy, Hansen:2024ffk}. Another well-known example is the Roper resonance $N(1440)$, which couples strongly to both $N\pi$ and $N\pi\pi$ channels and exhibits a nontrivial line shape. For a recent review of the experimental and theoretical status, see Ref.~\cite{Burkert:2017djo}.

In this section, we present several explicit examples of multi-hadron operators. Operators involving higher momentum shells or more constituent particles can be systematically constructed using the \texttt{OpTion} package introduced earlier. In Tab~\ref{tab:three-000} and~\ref{tab:three-moving}, we provide representative examples of three-body operators composed entirely of pseudoscalar mesons, both in the rest frame and moving frames.

\begin{table}[htbp]
\centering
\caption{A few examples of three-pseudoscalar operators in $O_h$ group. The notations are as in Tab.~\ref{tab:mm-000-A1+}.}
\addtolength{\tabcolsep}{-4pt}
\begin{tabular}{cc}
\toprule
irrep & operator \\
\midrule
\multirow{16}{*}{$A_1^+$} & {\small $- P_{1}(e_{y,z})P_{2}(e_{x,-z})P_{3}(e_{-x,-y}) + P_{1}(e_{y,-z})P_{2}(e_{x,z})P_{3}(e_{-x,-y}) + P_{1}(e_{x,z})P_{2}(e_{y,-z})P_{3}(e_{-x,-y})$} \\
& {\small $- P_{1}(e_{x,-z})P_{2}(e_{y,z})P_{3}(e_{-x,-y}) + P_{1}(e_{-y,z})P_{2}(e_{x,-z})P_{3}(e_{-x,y}) - P_{1}(e_{-y,-z})P_{2}(e_{x,z})P_{3}(e_{-x,y})$} \\
& {\small $- P_{1}(e_{x,z})P_{2}(e_{-y,-z})P_{3}(e_{-x,y}) + P_{1}(e_{x,-z})P_{2}(e_{-y,z})P_{3}(e_{-x,y}) + P_{1}(e_{y,z})P_{2}(e_{x,-y})P_{3}(e_{-x,-z})$} \\
& {\small $- P_{1}(e_{-y,z})P_{2}(e_{x,y})P_{3}(e_{-x,-z}) - P_{1}(e_{x,y})P_{2}(e_{-y,z})P_{3}(e_{-x,-z}) + P_{1}(e_{x,-y})P_{2}(e_{y,z})P_{3}(e_{-x,-z})$} \\
& {\small $- P_{1}(e_{y,-z})P_{2}(e_{x,-y})P_{3}(e_{-x,z}) + P_{1}(e_{-y,-z})P_{2}(e_{x,y})P_{3}(e_{-x,z}) + P_{1}(e_{x,y})P_{2}(e_{-y,-z})P_{3}(e_{-x,z})$} \\
& {\small $- P_{1}(e_{x,-y})P_{2}(e_{y,-z})P_{3}(e_{-x,z}) + P_{1}(e_{y,z})P_{2}(e_{-x,-z})P_{3}(e_{x,-y}) - P_{1}(e_{y,-z})P_{2}(e_{-x,z})P_{3}(e_{x,-y})$} \\
& {\small $- P_{1}(e_{-x,z})P_{2}(e_{y,-z})P_{3}(e_{x,-y}) + P_{1}(e_{-x,-z})P_{2}(e_{y,z})P_{3}(e_{x,-y}) - P_{1}(e_{-y,z})P_{2}(e_{-x,-z})P_{3}(e_{x,y})$} \\
& {\small $+ P_{1}(e_{-y,-z})P_{2}(e_{-x,z})P_{3}(e_{x,y}) + P_{1}(e_{-x,z})P_{2}(e_{-y,-z})P_{3}(e_{x,y}) - P_{1}(e_{-x,-z})P_{2}(e_{-y,z})P_{3}(e_{x,y})$} \\
& {\small $- P_{1}(e_{y,z})P_{2}(e_{-x,-y})P_{3}(e_{x,-z}) + P_{1}(e_{-y,z})P_{2}(e_{-x,y})P_{3}(e_{x,-z}) + P_{1}(e_{-x,y})P_{2}(e_{-y,z})P_{3}(e_{x,-z})$} \\
& {\small $- P_{1}(e_{-x,-y})P_{2}(e_{y,z})P_{3}(e_{x,-z}) + P_{1}(e_{y,-z})P_{2}(e_{-x,-y})P_{3}(e_{x,z}) - P_{1}(e_{-y,-z})P_{2}(e_{-x,y})P_{3}(e_{x,z})$} \\
& {\small $- P_{1}(e_{-x,y})P_{2}(e_{-y,-z})P_{3}(e_{x,z}) + P_{1}(e_{-x,-y})P_{2}(e_{y,-z})P_{3}(e_{x,z}) - P_{1}(e_{x,z})P_{2}(e_{-x,y})P_{3}(e_{-y,-z})$} \\
& {\small $+ P_{1}(e_{x,y})P_{2}(e_{-x,z})P_{3}(e_{-y,-z}) + P_{1}(e_{-x,z})P_{2}(e_{x,y})P_{3}(e_{-y,-z}) - P_{1}(e_{-x,y})P_{2}(e_{x,z})P_{3}(e_{-y,-z})$} \\
& {\small $P_{1}(e_{x,-z})P_{2}(e_{-x,y})P_{3}(e_{-y,z}) - P_{1}(e_{x,y})P_{2}(e_{-x,-z})P_{3}(e_{-y,z}) - P_{1}(e_{-x,-z})P_{2}(e_{x,y})P_{3}(e_{-y,z})$} \\
& {\small $+ P_{1}(e_{-x,y})P_{2}(e_{x,-z})P_{3}(e_{-y,z}) + P_{1}(e_{x,z})P_{2}(e_{-x,-y})P_{3}(e_{y,-z}) - P_{1}(e_{x,-y})P_{2}(e_{-x,z})P_{3}(e_{y,-z})$} \\
& {\small $- P_{1}(e_{-x,z})P_{2}(e_{x,-y})P_{3}(e_{y,-z}) + P_{1}(e_{-x,-y})P_{2}(e_{x,z})P_{3}(e_{y,-z}) - P_{1}(e_{x,-z})P_{2}(e_{-x,-y})P_{3}(e_{y,z})$} \\
& {\small $+ P_{1}(e_{x,-y})P_{2}(e_{-x,-z})P_{3}(e_{y,z}) + P_{1}(e_{-x,-z})P_{2}(e_{x,-y})P_{3}(e_{y,z}) - P_{1}(e_{-x,-y})P_{2}(e_{x,-z})P_{3}(e_{y,z})$} \\
\midrule
\multirow{1}{*}{$A_1^-$} & $P_{1}(0)P_{2}(0)P_{3}(0)$ \\
\midrule
\multirow{4}{*}{$E^+$} & $- P_{1}(e_{x,-y,z})P_{2}(e_{y})P_{3}(e_{-x,-z}) + P_{1}(e_{x,-y,-z})P_{2}(e_{y})P_{3}(e_{-x,z})$ \\
& $P_{1}(e_{-x,-y,z})P_{2}(e_{y})P_{3}(e_{x,-z}) - P_{1}(e_{-x,-y,-z})P_{2}(e_{y})P_{3}(e_{x,z})$ \\
& $P_{1}(e_{-x,y,z})P_{2}(e_{x})P_{3}(e_{-y,-z}) - P_{1}(e_{-x,y,-z})P_{2}(e_{x})P_{3}(e_{-y,z})$ \\
& $- P_{1}(e_{-x,-y,z})P_{2}(e_{x})P_{3}(e_{y,-z}) + P_{1}(e_{-x,-y,-z})P_{2}(e_{x})P_{3}(e_{y,z})$ \\
\midrule
\multirow{3}{*}{$E^-$} & $P_{1}(e_{x}) P_{2}(0) P_{3}(-e_{x}) + P_{1}(-e_{x}) P_{2}(0) P_{3}(e_{x})$ \\
& $P_{1}(e_{y}) P_{2}(0) P_{3}(-e_{y}) + P_{1}(-e_{y}) P_{2}(0) P_{3}(e_{y})$ \\
& $-2 P_{1}(e_{z}) P_{2}(0) P_{3}(-e_{z}) -2 P_{1}(-e_{z}) P_{2}(0) P_{3}(e_{z})$ \\
\midrule
\multirow{1}{*}{$T_1^+$} & $P_{1}(e_{x}) P_{2}(0) P_{3}(-e_{x}) - P_{1}(-e_{x}) P_{2}(0) P_{3}(e_{x})$ \\
\midrule
\multirow{4}{*}{$T_1^-$} & $P_{1}(e_{y,z}) P_{2}(-e_{z}) P_{3}(-e_{y}) - P_{1}(e_{y,-z}) P_{2}(e_{z}) P_{3}(-e_{y})$ \\
& $- P_{1}(e_{-y,z}) P_{2}(-e_{z}) P_{3}(e_{y}) + P_{1}(e_{-y,-z}) P_{2}(e_{z}) P_{3}(e_{y})$ \\
& $- P_{1}(e_{y,z}) P_{2}(-e_{y}) P_{3}(-e_{z}) + P_{1}(e_{-y,z}) P_{2}(e_{y}) P_{3}(-e_{z})$ \\
& $+ P_{1}(e_{y,-z}) P_{2}(-e_{y}) P_{3}(e_{z}) - P_{1}(e_{-y,-z}) P_{2}(e_{y}) P_{3}(e_{z})$ \\
\midrule
\multirow{4}{*}{$T_2^+$} & $P_{1}(e_{x,y}) P_{2}(0) P_{3}(e_{-x,-y}) + P_{1}(e_{x,-y}) P_{2}(0) P_{3}(e_{-x,y})$ \\
& $- P_{1}(e_{x,z}) P_{2}(0) P_{3}(e_{-x,-z}) - P_{1}(e_{x,-z}) P_{2}(0) P_{3}(e_{-x,z})$ \\
& $- P_{1}(e_{-x,y}) P_{2}(0) P_{3}(e_{x,-y}) - P_{1}(e_{-x,-y}) P_{2}(0) P_{3}(e_{x,y})$ \\
& $+ P_{1}(e_{-x,z}) P_{2}(0) P_{3}(e_{x,-z}) + P_{1}(e_{-x,-z}) P_{2}(0) P_{3}(e_{x,z})$ \\
\midrule
\multirow{2}{*}{$T_2^-$} & $P_{1}(e_{y,z}) P_{2}(0) P_{3}(e_{-y,-z}) - P_{1}(e_{y,-z}) P_{2}(0) P_{3}(e_{-y,z})$ \\
& $- P_{1}(e_{-y,z}) P_{2}(0) P_{3}(e_{y,-z}) + P_{1}(e_{-y,-z}) P_{2}(0) P_{3}(e_{y,z})$ \\
\bottomrule
\end{tabular}
\addtolength{\tabcolsep}{4pt}
\label{tab:three-000}
\end{table}

\begin{table}[htbp]
\centering
\caption{Examples of three-pseudoscalar operators in $C_{4v}$, $C_{4v}$, and $C_{3v}$ group. The notations are as in Tab.~\ref{tab:mm-000-A1+}.}
\addtolength{\tabcolsep}{6pt}
\begin{tabular}{ccc}
\toprule
group & irrep & operator \\
\midrule
\multirow{10}{*}{$C_{4v}$} & \multirow{4}{*}{$A_1$} & $P_{1}(e_{x,y}) P_{2}(-e_{y}) P_{3}(e_{-x,z}) - P_{1}(e_{x,-y}) P_{2}(e_{y}) P_{3}(e_{-x,z})$ \\
& & $- P_{1}(e_{-x,y}) P_{2}(-e_{y}) P_{3}(e_{x,z}) + P_{1}(e_{-x,-y}) P_{2}(e_{y}) P_{3}(e_{x,z})$ \\
& & $- P_{1}(e_{x,y}) P_{2}(-e_{x}) P_{3}(e_{-y,z}) + P_{1}(e_{-x,y}) P_{2}(e_{x}) P_{3}(e_{-y,z})$ \\
& & $+ P_{1}(e_{x,-y}) P_{2}(-e_{x}) P_{3}(e_{y,z}) - P_{1}(e_{-x,-y}) P_{2}(e_{x}) P_{3}(e_{y,z})$ \\
\cmidrule(lr){2-3}
& \multirow{1}{*}{$A_2$} & $P_{1}(0) P_{2}(0) P_{3}(e_{z})$ \\
\cmidrule(lr){2-3}
& \multirow{2}{*}{$B_1$} & $P_{1}(e_{x,y}) P_{2}(0) P_{3}(e_{-x,-y,z}) - P_{1}(e_{x,-y}) P_{2}(0) P_{3}(e_{-x,y,z})$ \\
& & $- P_{1}(e_{-x,y}) P_{2}(0) P_{3}(e_{x,-y,z}) + P_{1}(e_{-x,-y}) P_{2}(0) P_{3}(e_{x,y,z})$ \\
\cmidrule(lr){2-3}
& \multirow{2}{*}{$B_2$} & $P_{1}(e_{x}) P_{2}(e_{z}) P_{3}(-e_{x}) + P_{1}(-e_{x}) P_{2}(e_{z}) P_{3}(e_{x})$ \\
& & $- P_{1}(e_{y}) P_{2}(e_{z}) P_{3}(-e_{y}) - P_{1}(-e_{y}) P_{2}(e_{z}) P_{3}(e_{y})$ \\
\cmidrule(lr){2-3}
& \multirow{1}{*}{$E$} & $P_{1}(e_{x}) P_{2}(e_{z}) P_{3}(-e_{x}) - P_{1}(-e_{x}) P_{2}(e_{z}) P_{3}(e_{x})$ \\
\midrule
\multirow{5}{*}{$C_{2v}$} & \multirow{2}{*}{$A_1$} & $- P_{1}(e_{x}) P_{2}(e_{z}) P_{3}(e_{-x,y}) + P_{1}(e_{x}) P_{2}(e_{y}) P_{3}(e_{-x,z})$ \\
& & $+ P_{1}(-e_{x}) P_{2}(e_{z}) P_{3}(e_{x,y}) - P_{1}(-e_{x}) P_{2}(e_{y}) P_{3}(e_{x,z})$ \\
\cmidrule(lr){2-3}
& \multirow{1}{*}{$A_2$} & $P_{1}(0) P_{2}(e_{z}) P_{3}(e_{y}) + P_{1}(0) P_{2}(e_{y}) P_{3}(e_{z})$ \\
\cmidrule(lr){2-3}
& \multirow{1}{*}{$B_1$} & $P_{1}(e_{x}) P_{2}(0) P_{3}(e_{-x,y,z}) - P_{1}(-e_{x}) P_{2}(0) P_{3}(e_{x,y,z})$ \\
\cmidrule(lr){2-3}
& \multirow{1}{*}{$B_2$} & $P_{1}(0) P_{2}(e_{z}) P_{3}(e_{y}) - P_{1}(0) P_{2}(e_{y}) P_{3}(e_{z})$ \\
\midrule
\multirow{8}{*}{$C_{3v}$} & \multirow{3}{*}{$A_1$} & $P_{1}(e_{z}) P_{2}(e_{y}) P_{3}(e_{x}) - P_{1}(e_{y}) P_{2}(e_{z}) P_{3}(e_{x})$ \\
& & $- P_{1}(e_{z}) P_{2}(e_{x}) P_{3}(e_{y}) + P_{1}(e_{x}) P_{2}(e_{z}) P_{3}(e_{y})$ \\
& & $+ P_{1}(e_{y}) P_{2}(e_{x}) P_{3}(e_{z}) - P_{1}(e_{x}) P_{2}(e_{y}) P_{3}(e_{z})$ \\
\cmidrule(lr){2-3}
& \multirow{3}{*}{$A_2$} & $P_{1}(e_{z}) P_{2}(e_{y}) P_{3}(e_{x}) + P_{1}(e_{y}) P_{2}(e_{z}) P_{3}(e_{x})$ \\
& & $+ P_{1}(e_{z}) P_{2}(e_{x}) P_{3}(e_{y}) + P_{1}(e_{x}) P_{2}(e_{z}) P_{3}(e_{y})$ \\
& & $+ P_{1}(e_{y}) P_{2}(e_{x}) P_{3}(e_{z}) + P_{1}(e_{x}) P_{2}(e_{y}) P_{3}(e_{z})$ \\
\cmidrule(lr){2-3}
& \multirow{2}{*}{$E$} & $2 P_{1}(e_{z}) P_{2}(e_{y}) P_{3}(e_{x}) - P_{1}(e_{y}) P_{2}(e_{z}) P_{3}(e_{x})$ \\
& & $+ 2 P_{1}(e_{z}) P_{2}(e_{x}) P_{3}(e_{y}) - P_{1}(e_{x}) P_{2}(e_{z}) P_{3}(e_{y})$ \\
& & $- P_{1}(e_{y}) P_{2}(e_{x}) P_{3}(e_{z}) - P_{1}(e_{x}) P_{2}(e_{y}) P_{3}(e_{z})$ \\
\bottomrule
\end{tabular}
\addtolength{\tabcolsep}{-6pt}
\label{tab:three-moving}
\end{table}

In Tab~\ref{tab:four-000}, we list examples of four-body operators of distinguishable pseudoscalars at the rest frame. These operators serve as a foundation for future four-body spectroscopy calculations once the relevant quantization conditions are mature. For systems with even more particles, instead of giving a comprehensive list, we point out a special case that $\prod_i^N P_i(0)$, composed of $N$ pseudoscalar mesons at rest, transforms according to the $A_1^-$ irrep of the $O_h$ group when $N$ is odd, and the $A_1^+$ irrep when $N$ is even.

\begin{table}[htbp]
\centering
\caption{Examples of four-pseudoscalar operators in $O_h$ group. Only $A_1^+$ and $A_1^-$ irreps are listed. The notations are as in Tab.~\ref{tab:mm-000-A1+}.}
\addtolength{\tabcolsep}{6pt}
\begin{tabular}{cc}
\toprule
irrep & operator \\
\midrule
\multirow{3}{*}{$A_1^+$} & $P_{1}(0) P_{2}(0) P_{3}(0) P_{4}(0)$ \\
\cmidrule(lr){2-2}
& $P_{1}(e_{x}) P_{2}(0) P_{3}(0) P_{4}(-e_{x}) + P_{1}(-e_{x}) P_{2}(0) P_{3}(0) P_{4}(e_{x})$ \\
& $+ P_{1}(e_{y}) P_{2}(0) P_{3}(0) P_{4}(-e_{y}) + P_{1}(-e_{y}) P_{2}(0) P_{3}(0) P_{4}(e_{y})$ \\
& $+ P_{1}(e_{z}) P_{2}(0) P_{3}(0) P_{4}(-e_{z}) + P_{1}(-e_{z}) P_{2}(0) P_{3}(0) P_{4}(e_{z})$ \\
\midrule
\multirow{3}{*}{$E^+$} & $P_{1}(e_{x}) P_{2}(0) P_{3}(0) P_{4}(-e_{x}) + P_{1}(-e_{x}) P_{2}(0) P_{3}(0) P_{4}(e_{x})$ \\
& $+ P_{1}(e_{y}) P_{2}(0) P_{3}(0) P_{4}(-e_{y}) + P_{1}(-e_{y}) P_{2}(0) P_{3}(0) P_{4}(e_{y})$ \\
& $- 2 P_{1}(e_{z}) P_{2}(0) P_{3}(0) P_{4}(-e_{z}) - 2 P_{1}(-e_{z}) P_{2}(0) P_{3}(0) P_{4}(e_{z})$ \\
\midrule
\multirow{4}{*}{$T_1^+$} & $P_{1}(e_{y}) P_{2}(e_{z}) P_{3}(-e_{z}) P_{4}(-e_{y}) - P_{1}(e_{y}) P_{2}(-e_{z}) P_{3}(e_{z}) P_{4}(-e_{y})$ \\
& $- P_{1}(-e_{y}) P_{2}(e_{z}) P_{3}(-e_{z}) P_{4}(e_{y}) + P_{1}(-e_{y}) P_{2}(-e_{z}) P_{3}(e_{z}) P_{4}(e_{y})$ \\
& $- P_{1}(e_{z}) P_{2}(e_{y}) P_{3}(-e_{y}) P_{4}(-e_{z}) + P_{1}(e_{z}) P_{2}(-e_{y}) P_{3}(e_{y}) P_{4}(-e_{z})$ \\
& $+ P_{1}(-e_{z}) P_{2}(e_{y}) P_{3}(-e_{y}) P_{4}(e_{z}) - P_{1}(-e_{z}) P_{2}(-e_{y}) P_{3}(e_{y}) P_{4}(e_{z})$ \\
\midrule
\multirow{1}{*}{$T_1^-$} & $P_{1}(e_{x}) P_{2}(0) P_{3}(0) P_{4}(-e_{x}) - P_{1}(-e_{x}) P_{2}(0) P_{3}(0) P_{4}(e_{x})$ \\
\midrule
\multirow{4}{*}{$T_2^+$} & $P_{1}(e_{y}) P_{2}(e_{z}) P_{3}(-e_{z}) P_{4}(-e_{y}) - P_{1}(e_{y}) P_{2}(-e_{z}) P_{3}(e_{z}) P_{4}(-e_{y})$ \\
& $- P_{1}(-e_{y}) P_{2}(e_{z}) P_{3}(-e_{z}) P_{4}(-e_{y}) + P_{1}(-e_{y}) P_{2}(-e_{z}) P_{3}(e_{z}) P_{4}(e_{y})$ \\
& $+ P_{1}(e_{z}) P_{2}(e_{y}) P_{3}(-e_{y}) P_{4}(-e_{z}) - P_{1}(e_{z}) P_{2}(-e_{y}) P_{3}(e_{y}) P_{4}(-e_{z})$ \\
& $- P_{1}(-e_{z}) P_{2}(e_{y}) P_{3}(-e_{y}) P_{4}(e_{z}) + P_{1}(-e_{z}) P_{2}(-e_{y}) P_{3}(e_{y}) P_{4}(e_{z})$ \\
\midrule
\multirow{4}{*}{$T_2^-$} & $P_{1}(e_{x}) P_{2}(e_{y}) P_{3}(-e_{y}) P_{4}(-e_{x}) + P_{1}(e_{x}) P_{2}(-e_{y}) P_{3}(e_{y}) P_{4}(-e_{x})$ \\
& $- P_{1}(e_{x}) P_{2}(e_{z}) P_{3}(-e_{z}) P_{4}(-e_{x}) - P_{1}(e_{x}) P_{2}(-e_{z}) P_{3}(e_{z}) P_{4}(-e_{x})$ \\
& $- P_{1}(-e_{x}) P_{2}(e_{y}) P_{3}(-e_{y}) P_{4}(e_{x}) - P_{1}(-e_{x}) P_{2}(-e_{y}) P_{3}(e_{y}) P_{4}(e_{x})$ \\
& $+ P_{1}(-e_{x}) P_{2}(e_{z}) P_{3}(-e_{z}) P_{4}(e_{x}) + P_{1}(-e_{x}) P_{2}(-e_{z}) P_{3}(e_{z}) P_{4}(e_{x})$ \\
\bottomrule
\end{tabular}
\addtolength{\tabcolsep}{-6pt}
\label{tab:four-000}
\end{table}

\section{The projection onto internal quantum numbers}
\label{sec:int}
In this section, we consider internal symmetries, including isospin, $C$-parity, $G$-parity, and $P$-parity. Operators projected onto these internal symmetries typically provide better separation of states and are essential for identifying the desired bound states, virtual states, and resonances.

\subsection{Isospin}
In lattice simulations with degenerate up and down quarks, isospin symmetry is preserved, allowing us to further project the operators into total isospin and its third component, $I$ and $I_z$. This symmetry can be exploited to project interpolating operators onto definite isospin channels. As an illustrative example, we consider the pion, which forms an isospin triplet:
\begin{equation}
\begin{cases}
    | I=1, I_z=+1 \rangle &= -\bar{d} \gamma_5 u = - \pi^+, \\
    | I=1, I_z=0 \rangle &= \frac{1}{\sqrt{2}} (\bar{u} \gamma_5 u - \bar{d} \gamma_5 d) = \pi^0, \\
    | I=1, I_z=-1 \rangle &= \bar{u} \gamma_5 d = \pi^-.
\end{cases}
\end{equation}
The minus sign in front of $\pi^+$ is necessary to ensure that the antiquark doublet $(-\bar{d}, u)^T$ transforms in the same way as the quark doublet $(u, d)^T$ under $SU(2)$ transformations. See, for example, Ref.~\cite{Thomson:2013zua}.

For the two-particle system, by selecting the largest $I_z$ for each channel and applying the Clebsch-Gordon coefficients, the operators in flavor space should take the following forms:
\begin{equation}
\begin{cases}
    | \pi\pi \rangle_{I=2} &= \pi^+ \pi^+, \\
    | \pi\pi \rangle_{I=1} &= -\pi^+ \pi^0 + \pi^0 \pi^+, \\
    | \pi\pi \rangle_{I=0} &= -\frac{1}{\sqrt{3}} (\pi^+ \pi^- + \pi^- \pi^+ + \pi^0 \pi^0).
\end{cases}
\end{equation}

For the three-particle system, the isospin structure becomes more intricate. When adding three isospins, we have
\begin{equation}
1 \otimes 1 \otimes 1 = (0 \oplus 1 \oplus 2) \otimes 1 = 0 \oplus 1^3 \oplus 2^2 \oplus 3.
\end{equation}
The total isospin states $I=1$ and $I=2$ have degeneracies of $3$ and $2$, respectively. For the maximum ($I=3$) and minimum ($I=0$) isospins, such complexities do not arise. The form of the operators depends on the order in which the isospins are combined. In the following, we first combine the isospins of the first two particles, then add the third. Additionally, we consider only operators with $I_z = I$. A straightforward calculation yields
\begin{equation}
\begin{cases}
    | \pi\pi\pi \rangle_{I=3} & = -\pi^+\pi^+\pi^+, \\
    | \pi\pi\pi \rangle_{I=2,J_{12}=1} & = \frac{1}{\sqrt{2}} (\pi^+\pi^0\pi^+ - \pi^0\pi^+\pi^+), \\
    | \pi\pi\pi \rangle_{I=2,J_{12}=2} & = \frac{1}{\sqrt{6}} (2 \pi^+\pi^+\pi^0 - \pi^+\pi^0\pi^+ - \pi^0\pi^+\pi^+), \\
    | \pi\pi\pi \rangle_{I=1,J_{12}=0} & = \frac{1}{\sqrt{3}} ( \pi^+\pi^-\pi^+ + \pi^0\pi^0\pi^+ + \pi^-\pi^+\pi^+ ), \\
    | \pi\pi\pi \rangle_{I=1,J_{12}=1} & = \frac{1}{2} ( - \pi^+\pi^0\pi^0 + \pi^0\pi^+\pi^0 - \pi^+\pi^-\pi^+ + \pi^-\pi^+\pi^+ ), \\
    | \pi\pi\pi \rangle_{I=1,J_{12}=2} & = \frac{1}{\sqrt{60}} ( 6 \pi^+\pi^+\pi^- + 3 \pi^+\pi^0\pi^0 + 3 \pi^0\pi^+\pi^0 + \pi^+\pi^-\pi^+ - 2 \pi^0\pi^0\pi^+ + \pi^-\pi^+\pi^+), \\
    | \pi\pi\pi \rangle_{I=0} & = \frac{1}{\sqrt{6}} (-\pi^+\pi^0\pi^- + \pi^0\pi^+\pi^- + \pi^+\pi^-\pi^0 - \pi^-\pi^+\pi^0 - \pi^0\pi^-\pi^+ + \pi^-\pi^0\pi^+),
\end{cases}
\label{eq:pipipi}
\end{equation}
where $J_{12}$ represents the isospin of the first two particles. These operators have been employed in the study of $\pi\pi\pi$ scattering, at various isospin~\cite{Hansen:2020otl, Mai:2018djl, Blanton:2019vdk, Culver:2019vvu, Fischer:2020jzp, Mai:2021nul, Yan:2024gwp}.

Extending this approach to systems with higher isospin is straightforward. Additionally, for ensembles with different flavor symmetry groups--such as $SU(3)_f$--the approach remains largely unchanged, with the only modification being the use of Clebsch-Gordon coefficients appropriate to the relevant symmetry group.

To take full advantage of both spatial and internal symmetries, operator construction proceeds in two steps. First, spatially projected operators are built using the tables provided in Appendix~\ref{sec:append_list_mm} or generated using the \texttt{OpTion} package~\cite{github}. Next, appropriate internal quantum numbers are assigned to each constituent particle. These operators can then be further projected into definite $I, I_z$ using the routines described in this section.

\subsection{$C$-parity}
In certain channels, charge conjugation ($C$-parity) is a good quantum number and is preserved by the system and as a result, the interpolating operators employed in lattice QCD simulations or phenomenological studies should also respect this symmetry when applicable We denote the charge party transformation by $\mathcal{C}$ and the corresponding matrix representation as $C$. We start with the definitions and basic properties of the gamma matrices in Euclidean space (see, e.g., Ref.~\cite{Gattringer:2010zz}. The gamma matrices satisfy the anticommutation relation:
\begin{equation}
\{\gamma_\mu, \gamma_\nu \}=2 \delta_{\mu \nu} \mathbbm{1}, \,(\mu=1,2,3,4,5),
\end{equation}
where $\gamma_5=\gamma_1 \gamma_2 \gamma_3 \gamma_4$ and $C=\mathrm{i} \gamma_2 \gamma_4$. The gamma matrices transform under $C$ as:
\begin{equation}
\begin{aligned}
    C \mathbbm{1} C^{-1} &= \mathbbm{1}, \\
    C \gamma_5 C^{-1} &= \gamma_5, \\
    C \gamma_\mu C^{-1} &= -\gamma_\mu^T \quad, \\
    C \gamma_i\gamma_4 C^{-1} &= -(\gamma_i\gamma_4)^T, \\
    C \gamma_{\mu}\gamma_5 C^{-1} &= (\gamma_{\mu}\gamma_5)^T, \\
    C \gamma_i\gamma_j C^{-1} &= -(\gamma_i\gamma_j)^T, \, (i \neq j), \\
    C \gamma_4\gamma_5 C^{-1} &= (\gamma_4\gamma_5)^T, \\
\end{aligned}
\end{equation}
where $\mu \in [1,2,3,4]$ while $i \in [1,2,3]$.

Under charge conjugation, the quark fields transform as:
\begin{equation}
\begin{aligned}
& \psi(\vec{x},t) \xrightarrow{\mathcal{C}} C^{-1} \bar{\psi}(\vec{x},t)^T, \\
& \bar{\psi}(\vec{x},t) \xrightarrow{\mathcal{C}} -\psi(\vec{x},t)^T C,
\end{aligned}
\end{equation}
Using these transformation rules, one can deduce the behavior of quark bilinears under charge conjugation:
\begin{equation}
\begin{aligned}
    \psi_1 \mathbbm{1} \psi_2 &\xrightarrow{\mathcal{C}} + \psi_2 \mathbbm{1} \psi_1, \\
    \psi_1 \gamma_5 \psi_2 &\xrightarrow{\mathcal{C}} + \psi_2 \gamma_5 \psi_1, \\
    \psi_1 \gamma_\mu \psi_2 &\xrightarrow{\mathcal{C}} - \psi_2 \gamma_\mu \psi_1, \\
    \psi_1 \gamma_i\gamma_4 \psi_2 &\xrightarrow{\mathcal{C}} - \psi_2 \gamma_i\gamma_4 \psi_1, \\
    \psi_1 \gamma_{\mu}\gamma_5 \psi_2 &\xrightarrow{\mathcal{C}} + \psi_2 \gamma_{\mu}\gamma_5 \psi_1, \\
    \psi_1 \gamma_i\gamma_j \psi_2 &\xrightarrow{\mathcal{C}} - \psi_2 \gamma_i\gamma_j \psi_1 \\
    \psi_1 \gamma_4\gamma_5 \psi_2 &\xrightarrow{\mathcal{C}} + \psi_2 \gamma_4\gamma_5 \psi_1.
\end{aligned}
\end{equation}
Interpolating operators in channels where charge conjugation is a good symmetry should respect the corresponding transformation properties under $\mathcal{C}$. For example, in the study of $Z_c(3900)$ ($I^G(J^{PC}) = 1^+(1^{+-})$)~\cite{CLQCD:2019npr}, linear combinations of operators are chosen such that they project to $C=-1$.

\subsection{$G$-parity}
When we assume the isospin symmetry, the $C$-parity can be generalized to the $G$-parity for the isospin multiplet. The transformation property is similar to that of $\mathcal{C}$,
\begin{equation}
    \mathcal{G} = \mathcal{C} \operatorname{e}^{i\pi I_y},
\end{equation}
and the quantum number satisfies $G = C (-1)^I$. For details, refer to, e.g., Ref.~\cite{Christ:2019sah}. It can be shown that the exemplary operators in Eq.~\ref{eq:pipipi} have the correct $G$-parity when applying the transformation.

\subsection{$P$-parity}
In contrast to $C$-parity, which acts solely on the internal structure of hadrons, $P$-parity depends on both the intrinsic parity of the constituent particles and the orbital angular momentum between them. As a result, when operators are projected onto a specific irrep of the relevant little group, the resulting operator already possesses definite parity. This is consistent with the transformation properties discussed in Eq.~\ref{eq:trans}.

However, P-parity is a good quantum number only in the rest frame. In moving frames, as discussed in Refs.~\cite{Thomas:2011rh, Dudek:2012xn}, parity is generally not a symmetry of the finite-volume system because it does not commute with the reduced spatial symmetry group. As a result, energy levels associated with different parity channels can mix and complicate the analysis.

\texttt{OpTion} has already been extensively applied in several of our previous works and ongoing projects. For example, in our recent study of $\pi\pi\pi$ scattering~\cite{Yan:2024gwp}, effective mass plots generated using \texttt{OpTion} operators are shown in the supplementary material, and it is necessary to incorporate a complete basis of single-, two-, and three-body operators to achieve reliable and accurate determinations of finite-volume energy spectra. In studies of $D\pi$ scattering~\cite{Yan:2024yuq, Yan:2023gvq} and in an ongoing coupled-channel analysis of $D\pi-D\eta-D_s\bar{K}$, the complete operator basis has enabled precise determinations of the $D_0^*(2300)$ pole positions. Similar projection-based operator construction methods have been successfully used by the Hadron Spectrum Collaboration in state-of-the-art spectroscopy calculations, including extensive studies of charmonium resonances~\cite{Wilson:2023hzu, Wilson:2023anv}. In particular, Ref.~\cite{Dudek:2012xn} explicitly demonstrates the importance of using diverse operator bases to ensure correct extraction of finite-volume energies.

\section{Conclusions}
\label{sec:con}
In this work, we have systematically reviewed the structure of the cubic group and its relevant subgroups for lattice QCD applications, along with their irreps. We discussed projection techniques to construct operators and apply them to systems with arbitrary particle number, lattice momentum, and internal quantum numbers. This method has been implemented in the open-source package \texttt{OpTion}, which automates the construction of operators with definite symmetry properties and is available at \cite{github}.

We have provided explicit operator bases for one- and two-particle systems across all relevant lattice irreps, and presented examples for three- and four-particle constructions. The completeness of our operator sets has been cross-checked against the expected degrees of freedom from Dirac and spin representations. These operator dictionaries are intended to reduce redundant efforts in symmetry analysis and to provide a convenient starting point for future lattice studies.

Our construction framework enables investigations of increasingly complex systems, laying the foundation for the study of four-, five-, or even more-particle interactions in lattice QCD. \texttt{OpTion} has been a standard tool within the CLQCD Collaboration, significantly accelerating the process of obtaining energy spectra and reducing the barrier for students and newcomers to enter lattice spectroscopy.

We also briefly discussed the projection onto internal quantum numbers such as spin and isospin, ensuring that the constructed operators are suitable interpolators for states with well-defined physical quantum numbers. The resulting operators can be directly used in two-point and higher-point correlation functions for spectroscopy and scattering analyses. With the formalism of lattice spectroscopy maturing and computational resources steadily advancing, we look forward to an exciting era in which systematic and large-scale studies of multi-hadron systems become increasingly feasible.

\acknowledgments

H.Y. is grateful to the insightful discussions with M.~Garofalo, M.~Mai, S.~Prelovsek, J.~Suárez-Sucunza, J.~Wu, and C.~Urbach. Additional thanks are due to the early users of the \texttt{OpTion} package, particularly those from the CLQCD collaboration, for their valuable suggestions and feedback. H.Y. acknowledges support from NSFC under Grant No.~124B2096. H.Y., C.L., L.L., and Y.M. acknowledge support from NSFC under Grant No.~12293060, 12293061, 12293063, 12175279, 12305094, and 11935017. Y.M. thanks the support from the Young Elite Scientists Sponsorship Program by Henan Association for Science and Technology with Grant No. 2025HYTP003.

\appendix
\section{Code tutorials}
\label{sec:tutorials}
This section provides basic tutorials and explicit use examples of \texttt{OpTion} publicly available online~\cite{github}. We focus on the projection method discussed in the main text. For installation and detailed tutorials, see the \textit{Manual} folder inside the repository.

The function \verb|OneHadronOperatorAll[Ptot, Rep, RepRow, MaxND]| searches all possible one-hadron operators with the number of covariant derivatives less than \verb|MaxND| and returns the operator lists. As an example:
\begin{lstlisting}[language=Mathematica]
(* Particle settings *)
Ptot = {0, 0, 0};
Rep = "T1-";
RepRow = 3;
MaxND = 1;
(* Generate the one-hadron operators *)
Print/@OneHadronOperatorAll[Ptot, Rep, RepRow, MaxND];
\end{lstlisting}
Output: \\
\noindent $\small{\mathsf{V_z}}$ \\
\noindent $\small{\mathsf{D_z}}$ \\
\noindent $\small{\mathsf{A_y D_x - A_x D_y}}$ \\
These output corresponds to one section in Tab.~\ref{tab:one-000} in the main text, with row changed to $3$.

The function \verb|TwoHadronOperatorAll[Ptot, Rep, RepRow, MaxMom, Par1, Par2]| searches all possible two-hadron operators for the maximum momentum of individual particles less than \verb|MaxMom| and returns the operator lists:
\begin{lstlisting}[language=Mathematica]
(* Particle settings *)
Par1 = "P";
Par2 = "V";
Ptot = {0, 0, 0};
Rep = "T1-";
RepRow = 3;
MaxMom = 1;
(* Generate the two-hadron operators *)
Print/@TwoHadronOperatorAll[Ptot, Rep, RepRow, MaxMom, Par1, Par2];
\end{lstlisting}
Output: \\
\noindent $\small{\mathsf{-P_1\left[e_y\right] V_{2 x}\left[-e_y\right]+P_1\left[-e_y\right] V_{2 x}\left[e_y\right]+P_1\left[e_x\right] V_{2 y}\left[-e_x\right]-P_1\left[-e_x\right] V_{2 y}\left[e_x\right]}}$ \\

To construct operators from an arbitrary number of particles, one defines the argument \verb|ParTuple| to set the number of particles. As an example, here shows $\pi\pi\pi$ operators subduced into $T_1^+$ in the rest frame:
\begin{lstlisting}[language=Mathematica]
Rep = "T1+";
ParTuple = {"P", "P", "P"};
Print/@NHadronOperatorAll[Ptot, Rep, RepRow, MaxMom, ParTuple];
\end{lstlisting}
Output: \\
\noindent $\small{\mathsf{P_1\left[e_z\right] P_2[0] P_3\left[-e_z\right]-P_1\left[-e_z\right] P_2[0] P_3\left[e_z\right]}}$ \\
\noindent $\small{\mathsf{P_1\left[e_z\right] P_2\left[-e_z\right] P_3[0]-P_1\left[-e_z\right] P_2\left[e_z\right] P_3[0]}}$ \\
\noindent $\small{\mathsf{P_1[0] P_2\left[e_z\right] P_3\left[-e_z\right]-P_1[0] P_2\left[-e_z\right] P_3\left[e_z\right]}}$ \\

And these operators could be used to study the maximal isospin three-pion interactions.

\section{List of meson-meson operators}
\label{sec:append_list_mm}
Meson-meson operators with all $P,S,V,A$ structures and $|\vec{P}|^2 \leq 2$, continuing Sec.~\ref {sec:list_two} in the main text, are listed in Tab.~\ref{tab:mm-000-A1-}, \ref{tab:mm-000-E+} \ref{tab:mm-000-E-} \ref{tab:mm-000-T1+}, \ref{tab:mm-000-T1-}, \ref{tab:mm-000-T2+}, and \ref{tab:mm-000-T2-}.

\begin{table}[htbp]
\centering
\caption{Meson-meson operators in $A_1^-$ irrep of the cubic group $O_h$.}
\addtolength{\tabcolsep}{6pt}
\begin{tabular}{c}
\toprule
$P_1[0] S_2[0]$ \\
\midrule
$P_1[e_x] S_2[-e_x] + P_1[-e_x] S_2[e_x] + P_1[e_y] S_2[-e_y] + P_1[-e_y] S_2[e_y]$ \\
$+ P_1[e_z] S_2[-e_z] + P_1[-e_z] S_2[e_z]$ \\
\midrule
$P_1[e_x] V_{2,x}[-e_x] - P_1[-e_x] V_{2,x}[e_x] + P_1[e_y] V_{2,y}[-e_y] - P_1[-e_y] V_{2,y}[e_y]$ \\
$+ P_1[e_z] V_{2,z}[-e_z] - P_1[-e_z] V_{2,z}[e_z]$ \\
\midrule
$S_2[e_x] A_{1,x}[-e_x] - S_2[-e_x] A_{1,x}[e_x] + S_2[e_y] A_{1,y}[-e_y] - S_2[-e_y] A_{1,y}[e_y]$ \\
$+ S_2[e_z] A_{1,z}[-e_z] - S_2[-e_z] A_{1,z}[e_z]$ \\
\midrule
$V_{1,z}[e_y] V_{2,x}[-e_y] - V_{1,z}[-e_y] V_{2,x}[e_y] - V_{1,y}[e_z] V_{2,x}[-e_z] + V_{1,y}[-e_z] V_{2,x}[e_z]$ \\
\bottomrule
\end{tabular}
\addtolength{\tabcolsep}{-6pt}
\label{tab:mm-000-A1-}
\end{table}

\begin{table}[htbp]
\centering
\small
\caption{Same as Tab.~\ref{tab:mm-000-A1-}, with $G=O_h$, irrep $=E^+$.}
\addtolength{\tabcolsep}{6pt}
\begin{tabular}{c}
\toprule
$P_1[e_x] P_2[-e_x] + P_1[-e_x] P_2[e_x] + P_1[e_y] P_2[-e_y] + P_1[-e_y] P_2[e_y]$ \\ 
$-2 P_1[e_z] P_2[-e_z] -2 P_1[-e_z] P_2[e_z]$ \\ 
\midrule
$P_2[e_x] A_{1,x}[-e_x] - P_2[-e_x] A_{1,x}[e_x] + P_2[e_y] A_{1,y}[-e_y] - P_2[-e_y] A_{1,y}[e_y]$ \\ 
$-2 P_2[e_z] A_{1,z}[-e_z] + 2 P_2[-e_z] A_{1,z}[e_z]$ \\ 
\midrule
$S_1[e_x] S_2[-e_x] + S_1[-e_x] S_2[e_x] + S_1[e_y] S_2[-e_y] + S_1[-e_y] S_2[e_y]$ \\ 
$-2 S_1[e_z] S_2[-e_z] -2 S_1[-e_z] S_2[e_z]$ \\ 
\midrule
$S_1[e_x] V_{2,x}[-e_x] - S_1[-e_x] V_{2,x}[e_x] + S_1[e_y] V_{2,y}[-e_y] - S_1[-e_y] V_{2,y}[e_y]$ \\ 
$-2 S_1[e_z] V_{2,z}[-e_z] + 2 S_1[-e_z] V_{2,z}[e_z]$ \\
\bottomrule
\end{tabular}
\addtolength{\tabcolsep}{-6pt}
\label{tab:mm-000-E+}
\end{table}

\begin{table}[htbp]
\centering
\caption{Same as Tab.~\ref{tab:mm-000-A1-}, with $G=O_h$, irrep $=E^-$.}
\addtolength{\tabcolsep}{6pt}
\begin{tabular}{c}
\toprule
$P_1(e_x) S_2(-e_x) P_1(-e_x) S_2(e_x) P_1(e_y) S_2(-e_y) P_1(-e_y) S_2(e_y)$ \\
$-2 P_1(e_z) S_2(-e_z) -2 P_1(-e_z) S_2(e_z)$ \\
\midrule
$P_1(e_x) V_{2,x}(-e_x) -P_1(-e_x) V_{2,x}(e_x) P_1(e_y) V_{2,y}(-e_y) -P_1(-e_y) V_{2,y}(e_y)$ \\
$-2 P_1(e_z) V_{2,z}(-e_z) +2 P_1(-e_z) V_{2,z}(e_z)$ \\
\midrule
$S_2(e_x) A_{1,x}(-e_x) -S_2(-e_x) A_{1,x}(e_x) S_2(e_y) A_{1,y}(-e_y) -S_2(-e_y) A_{1,y}(e_y)$ \\
$-2 S_2(e_z) A_{1,z}(-e_z) +2 S_2(-e_z) A_{1,z}(e_z)$ \\
\midrule
$-V_{1,z}(e_y) V_{2,x}(-e_y) +V_{1,z}(-e_y) V_{2,x}(e_y) +V_{1,y}(e_z) V_{2,x}(-e_z) -V_{1,y}(-e_z) V_{2,x}(e_z)$ \\
$+V_{1,z}(e_x) V_{2,y}(-e_x) -V_{1,z}(-e_x) V_{2,y}(e_x) -V_{1,x}(e_z) V_{2,y}(-e_z) +V_{1,x}(-e_z) V_{2,y}(e_z)$ \\
\bottomrule
\end{tabular}
\addtolength{\tabcolsep}{-6pt}
\label{tab:mm-000-E-}
\end{table}

\begin{table}[htbp]
\centering
\caption{Same as Tab.~\ref{tab:mm-000-A1-}, with $G=O_h$, irrep $=T_1^+$.}
\addtolength{\tabcolsep}{6pt}
\begin{tabular}{c}
\toprule
$P_1(e_x) S_2(-e_x) -P_1(-e_x) S_2(e_x)$ \\
\midrule
$P_1(0) V_{2,x}(0)$ \\
\midrule
$P_1(e_x) V_{2,x}(-e_x) +P_1(-e_x) V_{2,x}(e_x)$ \\
\midrule
$P_1(e_y) V_{2,x}(-e_y) +P_1(-e_y) V_{2,x}(e_y) +P_1(e_z) V_{2,x}(-e_z) +P_1(-e_z) V_{2,x}(e_z)$ \\
\midrule
$P_2(e_z) A_{1,y}(-e_z) -P_2(-e_z) A_{1,y}(e_z) -P_2(e_y) A_{1,z}(-e_y) +P_2(-e_y) A_{1,z}(e_y)$ \\
\midrule
$S_1(e_z) V_{2,y}(-e_z) -S_1(-e_z) V_{2,y}(e_z) -S_1(e_y) V_{2,z}(-e_y) +S_1(-e_y) V_{2,z}(e_y)$ \\
\midrule
$S_2(0) A_{1,x}(0)$ \\
\midrule
$S_2(e_x) A_{1,x}(-e_x) +S_2(-e_x) A_{1,x}(e_x)$ \\
\bottomrule
\end{tabular}
\addtolength{\tabcolsep}{-6pt}
\label{tab:mm-000-T1+}
\end{table}

\begin{table}[htbp]
\centering
\caption{Same as Tab.~\ref{tab:mm-000-A1-}, with $G=O_h$, irrep $=T_1^-$.}
\addtolength{\tabcolsep}{6pt}
\begin{tabular}{c}
\toprule
$P_1(e_x) P_2(-e_x) -P_1(-e_x) P_2(e_x)$ \\
\midrule
$P_1(e_z) V_{2,y}(-e_z) -P_1(-e_z) V_{2,y}(e_z) -P_1(e_y) V_{2,z}(-e_y) +P_1(-e_y) V_{2,z}(e_y)$ \\
\midrule
$P_2(0) A_{1,x}(0)$ \\
\midrule
$P_2(e_x) A_{1,x}(-e_x) +P_2(-e_x) A_{1,x}(e_x)$ \\
\midrule
$P_2(e_y) A_{1,x}(-e_y) +P_2(-e_y) A_{1,x}(e_y) +P_2(e_z) A_{1,x}(-e_z) +P_2(-e_z) A_{1,x}(e_z)$ \\
\midrule
$S_1(e_x) S_2(-e_x) -S_1(-e_x) S_2(e_x)$ \\
\midrule
$S_1(0) V_{2,x}(0)$ \\
\midrule
$S_1(e_x) V_{2,x}(-e_x) +S_1(-e_x) V_{2,x}(e_x)$ \\
\bottomrule
\end{tabular}
\addtolength{\tabcolsep}{-6pt}
\label{tab:mm-000-T1-}
\end{table}

\begin{table}[htbp]
\centering
\caption{Same as Tab.~\ref{tab:mm-000-A1-}, with $G=O_h$, irrep $=T_2^+$.}
\addtolength{\tabcolsep}{6pt}
\begin{tabular}{c}
\toprule
$P_1(e_y) V_{2,x}(-e_y) + P_1(-e_y) V_{2,x}(e_y) - P_1(e_z) V_{2,x}(-e_z) - P_1(-e_z) V_{2,x}(e_z)$ \\
\midrule
$P_2(e_z) A_{1,y}(-e_z) - P_2(-e_z) A_{1,y}(e_z) + P_2(e_y) A_{1,z}(-e_y) - P_2(-e_y) A_{1,z}(e_y)$ \\
\midrule
$S_1(e_z) V_{2,y}(-e_z) - S_1(-e_z) V_{2,y}(e_z) + S_1(e_y) V_{2,z}(-e_y) - S_1(-e_y) V_{2,z}(e_y)$ \\
\midrule
$S_2(e_y) A_{1,x}(-e_y) + S_2(-e_y) A_{1,x}(e_y) - S_2(e_z) A_{1,x}(-e_z) - S_2(-e_z) A_{1,x}(e_z)$ \\
\midrule
$V_{1,z}(0) V_{2,y}(0) + V_{1,y}(0) V_{2,z}(0)$ \\
\midrule
$V_{1,z}(e_x) V_{2,y}(-e_x) + V_{1,z}(-e_x) V_{2,y}(e_x) + V_{1,y}(e_x) V_{2,z}(-e_x) + V_{1,y}(-e_x) V_{2,z}(e_x)$ \\
\midrule
$V_{1,z}(e_z) V_{2,y}(-e_z) + V_{1,z}(-e_z) V_{2,y}(e_z) + V_{1,y}(e_y) V_{2,z}(-e_y) + V_{1,y}(-e_y) V_{2,z}(e_y)$ \\
\midrule
$V_{1,z}(e_y) V_{2,y}(-e_y) + V_{1,z}(-e_y) V_{2,y}(e_y) + V_{1,y}(e_z) V_{2,z}(-e_z) + V_{1,y}(-e_z) V_{2,z}(e_z)$ \\
\bottomrule
\end{tabular}
\addtolength{\tabcolsep}{-6pt}
\label{tab:mm-000-T2+}
\end{table}

\begin{table}[htbp]
\centering
\caption{Same as Tab.~\ref{tab:mm-000-A1-}, with $G=O_h$, irrep $=T_2^-$.}
\addtolength{\tabcolsep}{6pt}
\begin{tabular}{c}
\toprule
$P_1(e_z) V_{2,y}(-e_z) -P_1(-e_z) V_{2,y}(e_z) +P_1(e_y) V_{2,z}(-e_y) -P_1(-e_y) V_{2,z}(e_y)$ \\
\midrule
$P_2(e_y) A_{1,x}(-e_y) +P_2(-e_y) A_{1,x}(e_y) -P_2(e_z) A_{1,x}(-e_z) -P_2(-e_z) A_{1,x}(e_z)$ \\
\midrule
$S_1(e_y) V_{2,x}(-e_y) +S_1(-e_y) V_{2,x}(e_y) -S_1(e_z) V_{2,x}(-e_z) -S_1(-e_z) V_{2,x}(e_z)$ \\
\midrule
$S_2(e_z) A_{1,y}(-e_z) -S_2(-e_z) A_{1,y}(e_z) +S_2(e_y) A_{1,z}(-e_y) -S_2(-e_y) A_{1,z}(e_y)$ \\
\midrule
$V_{1,y}(e_x) V_{2,y}(-e_x) -V_{1,y}(-e_x) V_{2,y}(e_x) -V_{1,z}(e_x) V_{2,z}(-e_x) +V_{1,z}(-e_x) V_{2,z}(e_x)$ \\
\midrule
$V_{1,y}(e_y) V_{2,x}(-e_y) -V_{1,y}(-e_y) V_{2,x}(e_y) -V_{1,z}(e_z) V_{2,x}(-e_z) +V_{1,z}(-e_z) V_{2,x}(e_z)$ \\
$+V_{1,x}(e_y) V_{2,y}(-e_y) -V_{1,x}(-e_y) V_{2,y}(e_y) -V_{1,x}(e_z) V_{2,z}(-e_z) +V_{1,x}(-e_z) V_{2,z}(e_z)$ \\
\midrule
$V_{1,y}(e_y) V_{2,x}(-e_y) -V_{1,y}(-e_y) V_{2,x}(e_y) -V_{1,z}(e_z) V_{2,x}(-e_z) +V_{1,z}(-e_z) V_{2,x}(e_z)$ \\
\bottomrule
\end{tabular}
\addtolength{\tabcolsep}{-6pt}
\label{tab:mm-000-T2-}
\end{table}

\begin{table}[htbp]
\centering
\caption{Same as Tab.~\ref{tab:mm-000-A1-}, with $\vec{P}=[0,0,n]$, irrep $=A_1$.}
\addtolength{\tabcolsep}{6pt}
\begin{tabular}{c}
\toprule
$P_1(0) P_2(e_z)$ \\
\midrule
$P_1(e_z) P_2(0)$ \\
\midrule
$P_2(e_z) A_{1,z}(0)$ \\
\midrule
$P_2(0) A_{1,z}(e_z)$ \\
\midrule
$S_1(0) S_2(e_z)$ \\
\midrule
$S_1(e_z) S_2(0)$ \\
\midrule
$S_1(0) V_{2,z}(e_z)$ \\
\midrule
$S_1(e_z) V_{2,z}(0)$ \\
\midrule
$V_{1,x}(0) V_{2,x}(e_z) + V_{1,y}(0) V_{2,y}(e_z)$ \\
\midrule
$V_{1,z}(0) V_{2,z}(e_z)$ \\
\midrule
$V_{1,x}(e_z) V_{2,x}(0) + V_{1,y}(e_z) V_{2,y}(0)$ \\
\midrule
$V_{1,z}(e_z) V_{2,z}(0)$ \\
\midrule
$-A_{1,y}(0) V_{2,x}(e_z) + A_{1,x}(0) V_{2,y}(e_z)$ \\
\midrule
$-A_{1,y}(e_z) V_{2,x}(0) + A_{1,x}(e_z) V_{2,y}(0)$ \\
\midrule
$A_{1,x}(0) A_{2,x}(e_z) + A_{1,y}(0) A_{2,y}(e_z)$ \\
\midrule
$A_{1,z}(0) A_{2,z}(e_z)$ \\
\midrule
$A_{1,x}(e_z) A_{2,x}(0) + A_{1,y}(e_z) A_{2,y}(0)$ \\
\midrule
$A_{1,z}(e_z) A_{2,z}(0)$ \\
\bottomrule
\end{tabular}
\addtolength{\tabcolsep}{-6pt}
\label{tab:mm-00n-A1}
\end{table}

\begin{table}[htbp]
\centering
\caption{Same as Tab.~\ref{tab:mm-000-A1-}, with $\vec{P}=[0,0,n]$, irrep $=A_2$.}
\addtolength{\tabcolsep}{6pt}
\begin{tabular}{c}
\toprule
$P_1(0) S_2(e_z)$ \\
\midrule
$P_1(e_z) S_2(0)$ \\
\midrule
$P_1(0) V_{2,z}(e_z) $ \\
\midrule
$P_1(e_z) V_{2,z}(0)$ \\
\midrule
$S_2(e_z) A_{1,z}(0)$ \\
\midrule
$S_2(0) A_{1,z}(e_z)$ \\
\midrule
$-V_{1,y}(0) V_{2,x}(e_z) + V_{1,x}(0) V_{2,y}(e_z)$ \\
\midrule
$-V_{1,y}(e_z) V_{2,x}(0) + V_{1,x}(e_z) V_{2,y}(0)$ \\
\midrule
$A_{1,x}(0) V_{2,x}(e_z) + A_{1,y}(0) V_{2,y}(e_z)$ \\
\midrule
$A_{1,z}(0) V_{2,z}(e_z)$ \\
\midrule
$A_{1,x}(e_z) V_{2,x}(0) + A_{1,y}(e_z) V_{2,y}(0)$ \\
\midrule
$A_{1,z}(e_z) V_{2,z}(0)$ \\
\midrule
$-A_{1,y}(0) A_{2,x}(e_z) + A_{1,x}(0) A_{2,y}(e_z)$ \\
\midrule
$-A_{1,y}(e_z) A_{2,x}(0) + A_{1,x}(e_z) A_{2,y}(0)$ \\
\bottomrule
\end{tabular}
\addtolength{\tabcolsep}{-6pt}
\label{tab:mm-00n-A2}
\end{table}

\begin{table}[htbp]
\centering
\caption{Same as Tab.~\ref{tab:mm-000-A1-}, with $\vec{P}=[0,0,n]$, irrep $=B_1$.}
\addtolength{\tabcolsep}{6pt}
\begin{tabular}{c}
\toprule
$V_{1,x}(0) V_{2,x}(e_z) -V_{1,y}(0) V_{2,y}(e_z)$ \\
\midrule
$V_{1,x}(e_z) V_{2,x}(0) -V_{1,y}(e_z) V_{2,y}(0)$ \\
\midrule
$A_{1,y}(0) V_{2,x}(e_z) +A_{1,x}(0) V_{2,y}(e_z)$ \\
\midrule
$A_{1,y}(e_z) V_{2,x}(0) +A_{1,x}(e_z) V_{2,y}(0)$ \\
\midrule
$A_{1,x}(0) A_{2,x}(e_z) -A_{1,y}(0) A_{2,y}(e_z)$ \\
\midrule
$A_{1,x}(e_z) A_{2,x}(0) -A_{1,y}(e_z) A_{2,y}(0)$ \\
\bottomrule
\end{tabular}
\addtolength{\tabcolsep}{-6pt}
\label{tab:mm-00n-B1}
\end{table}

\begin{table}[htbp]
\centering
\caption{Same as Tab.~\ref{tab:mm-000-A1-}, with $\vec{P}=[0,0,n]$, irrep $=B_2$.}
\addtolength{\tabcolsep}{6pt}
\begin{tabular}{c}
\toprule
$V_{1,y}(0) V_{2,x}(e_z) +V_{1,x}(0) V_{2,y}(e_z)$ \\
\midrule
$V_{1,y}(e_z) V_{2,x}(0) +V_{1,x}(e_z) V_{2,y}(0)$ \\
\midrule
$A_{1,x}(0) V_{2,x}(e_z) -A_{1,y}(0) V_{2,y}(e_z)$ \\
\midrule
$A_{1,x}(e_z) V_{2,x}(0) -A_{1,y}(e_z) V_{2,y}(0)$ \\
\midrule
$A_{1,y}(0) A_{2,x}(e_z) +A_{1,x}(0) A_{2,y}(e_z)$ \\
\midrule
$A_{1,y}(e_z) A_{2,x}(0) +A_{1,x}(e_z) A_{2,y}(0)$ \\
\bottomrule
\end{tabular}
\addtolength{\tabcolsep}{-6pt}
\label{tab:mm-00n-B2}
\end{table}

\begin{table}[htbp]
\centering
\caption{Same as Tab.~\ref{tab:mm-000-A1-}, with $\vec{P}=[0,0,n]$, irrep $=E$.}
\addtolength{\tabcolsep}{6pt}
\begin{tabular}{c}
\toprule
$P_1(0) V_{2,x}(e_z)$ \\
\midrule
$P_1(e_z) V_{2,x}(0)$ \\
\midrule
$P_2(e_z) A_{1,y}(0)$ \\
\midrule
$P_2(0) A_{1,y}(e_z)$ \\
\midrule
$S_1(0) V_{2,y}(e_z)$ \\
\midrule
$S_1(e_z) V_{2,y}(0)$ \\
\midrule
$S_2(e_z) A_{1,x}(0)$ \\
\midrule
$S_2(0) A_{1,x}(e_z)$ \\
\midrule
$V_{1,y}(0) V_{2,z}(e_z)$ \\
\midrule
$V_{1,z}(0) V_{2,y}(e_z)$ \\
\midrule
$V_{1,y}(e_z) V_{2,z}(0)$ \\
\midrule
$V_{1,z}(e_z) V_{2,y}(0)$ \\
\midrule
$A_{1,x}(0) V_{2,z}(e_z)$ \\
\midrule
$A_{1,z}(0) V_{2,x}(e_z)$ \\
\midrule
$A_{1,x}(e_z) V_{2,z}(0)$ \\
\midrule
$A_{1,z}(e_z) V_{2,x}(0)$ \\
\midrule
$A_{1,y}(0) A_{2,z}(e_z)$ \\
\midrule
$A_{1,z}(0) A_{2,y}(e_z)$ \\
\midrule
$A_{1,y}(e_z) A_{2,z}(0)$ \\
\midrule
$A_{1,z}(e_z) A_{2,y}(0)$ \\
\midrule
$A_{1,x}(e_z) V_{2,x}(0) -A_{1,y}(e_z) V_{2,y}(0)$ \\
\midrule
$A_{1,y}(0) A_{2,x}(e_z) +A_{1,x}(0) A_{2,y}(e_z)$ \\
\midrule
$A_{1,y}(e_z) A_{2,x}(0) +A_{1,x}(e_z) A_{2,y}(0)$ \\
\bottomrule
\end{tabular}
\addtolength{\tabcolsep}{-6pt}
\label{tab:mm-00n-E}
\end{table}

\begin{table}[htbp]
\centering
\caption{Same as Tab.~\ref{tab:mm-000-A1-}, with $\vec{P}=[0,n,n]$, irrep $=A_1$.}
\addtolength{\tabcolsep}{6pt}
\begin{tabular}{c}
\toprule
$P_1(e_z) P_2(e_y) +P_1(e_y) P_2(e_z)$ \\
\midrule
$P_1(e_z) V_{2,x}(e_y) -P_1(e_y) V_{2,x}(e_z)$ \\
\midrule
$P_2(e_y) A_{1,y}(e_z) +P_2(e_z) A_{1,z}(e_y)$ \\
\midrule
$P_2(e_z) A_{1,y}(e_y) +P_2(e_y) A_{1,z}(e_z)$ \\
\midrule
$S_1(e_z) S_2(e_y) +S_1(e_y) S_2(e_z)$ \\
\midrule
$S_1(e_z) V_{2,y}(e_y) +S_1(e_y) V_{2,z}(e_z)$ \\
\midrule
$S_1(e_y) V_{2,y}(e_z) +S_1(e_z) V_{2,z}(e_y)$ \\
\midrule
$-S_2(e_z) A_{1,x}(e_y) +S_2(e_y) A_{1,x}(e_z)$ \\
\midrule
$V_{1,x}(e_z) V_{2,x}(e_y) +V_{1,x}(e_y) V_{2,x}(e_z) -V_{1,y}(e_z) V_{2,y}(e_y) -V_{1,z}(e_y) V_{2,z}(e_z)$ \\
\midrule
$V_{1,z}(e_y) V_{2,y}(e_z) +V_{1,y}(e_z) V_{2,z}(e_y)$ \\
\midrule
$-V_{1,x}(e_z) V_{2,x}(e_y) -V_{1,x}(e_y) V_{2,x}(e_z) -V_{1,y}(e_z) V_{2,y}(e_y) -V_{1,z}(e_y) V_{2,z}(e_z)$ \\
\midrule
$V_{1,z}(e_z) V_{2,y}(e_y) +V_{1,y}(e_y) V_{2,z}(e_z)$ \\
\midrule
$V_{1,y}(e_y) V_{2,y}(e_z) +V_{1,z}(e_z) V_{2,z}(e_y)$ \\
\midrule
$A_{1,y}(e_z) V_{2,x}(e_y) -A_{1,z}(e_y) V_{2,x}(e_z) +A_{1,x}(e_z) V_{2,y}(e_y) -A_{1,x}(e_y) V_{2,z}(e_z)$ \\
\midrule
$A_{1,x}(e_y) V_{2,y}(e_z) -A_{1,x}(e_z) V_{2,z}(e_y)$ \\
\midrule
$A_{1,y}(e_z) V_{2,x}(e_y) -A_{1,z}(e_y) V_{2,x}(e_z) -A_{1,x}(e_z) V_{2,y}(e_y) +A_{1,x}(e_y) V_{2,z}(e_z)$ \\
\midrule
$A_{1,z}(e_z) V_{2,x}(e_y) -A_{1,y}(e_y) V_{2,x}(e_z)$ \\
\midrule
$A_{1,x}(e_z) A_{2,x}(e_y) +A_{1,x}(e_y) A_{2,x}(e_z) -A_{1,y}(e_z) A_{2,y}(e_y) -A_{1,z}(e_y) A_{2,z}(e_z)$ \\
\midrule
$A_{1,z}(e_y) A_{2,y}(e_z) +A_{1,y}(e_z) A_{2,z}(e_y)$ \\
\midrule
$-A_{1,x}(e_z) A_{2,x}(e_y) -A_{1,x}(e_y) A_{2,x}(e_z) -A_{1,y}(e_z) A_{2,y}(e_y) -A_{1,z}(e_y) A_{2,z}(e_z)$ \\
\midrule
$A_{1,z}(e_z) A_{2,y}(e_y) +A_{1,y}(e_y) A_{2,z}(e_z)$ \\
\midrule
$A_{1,y}(e_y) A_{2,y}(e_z) +A_{1,z}(e_z) A_{2,z}(e_y)$ \\
\bottomrule
\end{tabular}
\addtolength{\tabcolsep}{-6pt}
\label{tab:mm-0nn-A1}
\end{table}

\begin{table}[htbp]
\centering
\caption{Same as Tab.~\ref{tab:mm-000-A1-}, with $\vec{P}=[0,n,n]$, irrep $=A_2$.}
\addtolength{\tabcolsep}{6pt}
\begin{tabular}{c}
\toprule
$P_1(e_z) S_2(e_y) +P_1(e_y) S_2(e_z)$ \\
\midrule
$P_1(e_z) V_{2,y}(e_y) +P_1(e_y) V_{2,z}(e_z)$ \\
\midrule
$P_1(e_y) V_{2,y}(e_z) +P_1(e_z) V_{2,z}(e_y)$ \\
\midrule
$-P_2(e_z) A_{1,x}(e_y) +P_2(e_y) A_{1,x}(e_z)$ \\
\midrule
$S_1(e_z) V_{2,x}(e_y) -S_1(e_y) V_{2,x}(e_z)$ \\
\midrule
$S_2(e_y) A_{1,y}(e_z) +S_2(e_z) A_{1,z}(e_y)$ \\
\midrule
$S_2(e_z) A_{1,y}(e_y) +S_2(e_y) A_{1,z}(e_z)$ \\
\midrule
$V_{1,y}(e_z) V_{2,x}(e_y) -V_{1,z}(e_y) V_{2,x}(e_z)$ \\
$+V_{1,x}(e_z) V_{2,y}(e_y) -V_{1,x}(e_y) V_{2,z}(e_z)$ \\
\midrule
$V_{1,x}(e_y) V_{2,y}(e_z) -V_{1,x}(e_z) V_{2,z}(e_y)$ \\
\midrule
$V_{1,y}(e_z) V_{2,x}(e_y) -V_{1,z}(e_y) V_{2,x}(e_z)$ \\
$-V_{1,x}(e_z) V_{2,y}(e_y) +V_{1,x}(e_y) V_{2,z}(e_z)$ \\
\midrule
$V_{1,z}(e_z) V_{2,x}(e_y) -V_{1,y}(e_y) V_{2,x}(e_z)$ \\
\midrule
$A_{1,x}(e_z) V_{2,x}(e_y) +A_{1,x}(e_y) V_{2,x}(e_z)$ \\
$-A_{1,y}(e_z) V_{2,y}(e_y) -A_{1,z}(e_y) V_{2,z}(e_z)$ \\
\midrule
$A_{1,z}(e_y) V_{2,y}(e_z) +A_{1,y}(e_z) V_{2,z}(e_y)$ \\
\midrule
$-A_{1,x}(e_z) V_{2,x}(e_y) -A_{1,x}(e_y) V_{2,x}(e_z)$ \\
$-A_{1,y}(e_z) V_{2,y}(e_y) -A_{1,z}(e_y) V_{2,z}(e_z)$ \\
\midrule
$A_{1,z}(e_z) V_{2,y}(e_y) +A_{1,y}(e_y) V_{2,z}(e_z)$ \\
\midrule
$A_{1,y}(e_y) V_{2,y}(e_z) +A_{1,z}(e_z) V_{2,z}(e_y)$ \\
\midrule
$A_{1,y}(e_z) A_{2,x}(e_y) -A_{1,z}(e_y) A_{2,x}(e_z)$ \\
$+A_{1,x}(e_z) A_{2,y}(e_y) -A_{1,x}(e_y) A_{2,z}(e_z)$ \\
\midrule
$A_{1,x}(e_y) A_{2,y}(e_z) -A_{1,x}(e_z) A_{2,z}(e_y)$ \\
\midrule
$A_{1,y}(e_z) A_{2,x}(e_y) -A_{1,z}(e_y) A_{2,x}(e_z)$ \\
$-A_{1,x}(e_z) A_{2,y}(e_y) +A_{1,x}(e_y) A_{2,z}(e_z)$ \\
\midrule
$A_{1,z}(e_z) A_{2,x}(e_y) -A_{1,y}(e_y) A_{2,x}(e_z)$ \\
\bottomrule
\end{tabular}
\addtolength{\tabcolsep}{-6pt}
\label{tab:mm-0nn-A2}
\end{table}

\begin{table}[htbp]
\centering
\caption{Same as Tab.~\ref{tab:mm-000-A1-}, with $\vec{P}=[0,n,n]$, irrep $=B_1$.}
\addtolength{\tabcolsep}{6pt}
\begin{tabular}{c}
\toprule
$P_1(e_z) P_2(e_y) -P_1(e_y) P_2(e_z)$ \\
\midrule
$P_1(e_z) V_{2,x}(e_y) +P_1(e_y) V_{2,x}(e_z)$ \\
\midrule
$P_2(e_y) A_{1,y}(e_z) -P_2(e_z) A_{1,z}(e_y)$ \\
\midrule
$P_2(e_z) A_{1,y}(e_y) -P_2(e_y) A_{1,z}(e_z)$ \\
\midrule
$S_1(e_z) S_2(e_y) -S_1(e_y) S_2(e_z)$ \\
\midrule
$S_1(e_z) V_{2,y}(e_y) -S_1(e_y) V_{2,z}(e_z)$ \\
\midrule
$S_1(e_y) V_{2,y}(e_z) -S_1(e_z) V_{2,z}(e_y)$ \\
\midrule
$S_2(e_z) A_{1,x}(e_y) +S_2(e_y) A_{1,x}(e_z)$ \\
\midrule
$V_{1,x}(e_z) V_{2,x}(e_y) -V_{1,x}(e_y) V_{2,x}(e_z)$ \\
$-V_{1,y}(e_z) V_{2,y}(e_y) +V_{1,z}(e_y) V_{2,z}(e_z)$ \\
\midrule
$V_{1,z}(e_y) V_{2,y}(e_z) -V_{1,y}(e_z) V_{2,z}(e_y)$ \\
\midrule
$-V_{1,x}(e_z) V_{2,x}(e_y) +V_{1,x}(e_y) V_{2,x}(e_z)$ \\
$-V_{1,y}(e_z) V_{2,y}(e_y) +V_{1,z}(e_y) V_{2,z}(e_z)$ \\
\midrule
$V_{1,z}(e_z) V_{2,y}(e_y) -V_{1,y}(e_y) V_{2,z}(e_z)$ \\
\midrule
$V_{1,y}(e_y) V_{2,y}(e_z) -V_{1,z}(e_z) V_{2,z}(e_y)$ \\
\midrule
$A_{1,y}(e_z) V_{2,x}(e_y) +A_{1,z}(e_y) V_{2,x}(e_z)$ \\
$+A_{1,x}(e_z) V_{2,y}(e_y) +A_{1,x}(e_y) V_{2,z}(e_z)$ \\
\midrule
$A_{1,x}(e_y) V_{2,y}(e_z) +A_{1,x}(e_z) V_{2,z}(e_y)$ \\
\midrule
$A_{1,y}(e_z) V_{2,x}(e_y) +A_{1,z}(e_y) V_{2,x}(e_z)$ \\
$-A_{1,x}(e_z) V_{2,y}(e_y) -A_{1,x}(e_y) V_{2,z}(e_z)$ \\
\midrule
$A_{1,z}(e_z) V_{2,x}(e_y) +A_{1,y}(e_y) V_{2,x}(e_z)$ \\
\midrule
$A_{1,x}(e_z) A_{2,x}(e_y) -A_{1,x}(e_y) A_{2,x}(e_z)$ \\
$-A_{1,y}(e_z) A_{2,y}(e_y) +A_{1,z}(e_y) A_{2,z}(e_z)$ \\
\midrule
$A_{1,z}(e_y) A_{2,y}(e_z) -A_{1,y}(e_z) A_{2,z}(e_y)$ \\
\midrule
$-A_{1,x}(e_z) A_{2,x}(e_y) +A_{1,x}(e_y) A_{2,x}(e_z)$ \\
$-A_{1,y}(e_z) A_{2,y}(e_y) +A_{1,z}(e_y) A_{2,z}(e_z)$ \\
\midrule
$A_{1,z}(e_z) A_{2,y}(e_y) -A_{1,y}(e_y) A_{2,z}(e_z)$ \\
\midrule
$A_{1,y}(e_y) A_{2,y}(e_z) -A_{1,z}(e_z) A_{2,z}(e_y)$ \\
\bottomrule
\end{tabular}
\addtolength{\tabcolsep}{-6pt}
\label{tab:mm-0nn-B1}
\end{table}

\begin{table}[htbp]
\centering
\caption{Same as Tab.~\ref{tab:mm-000-A1-}, with $\vec{P}=[0,n,n]$, irrep $=B_2$.}
\addtolength{\tabcolsep}{6pt}
\begin{tabular}{c}
\toprule
$P_1(e_z) S_2(e_y) -P_1(e_y) S_2(e_z)$ \\
\midrule
$P_1(e_z) V_{2,y}(e_y) -P_1(e_y) V_{2,z}(e_z)$ \\
\midrule
$P_1(e_y) V_{2,y}(e_z) -P_1(e_z) V_{2,z}(e_y)$ \\
\midrule
$P_2(e_z) A_{1,x}(e_y) +P_2(e_y) A_{1,x}(e_z)$ \\
\midrule
$S_1(e_z) V_{2,x}(e_y) +S_1(e_y) V_{2,x}(e_z)$ \\
\midrule
$S_2(e_y) A_{1,y}(e_z) -S_2(e_z) A_{1,z}(e_y)$ \\
\midrule
$S_2(e_z) A_{1,y}(e_y) -S_2(e_y) A_{1,z}(e_z)$ \\
\midrule
$V_{1,y}(e_z) V_{2,x}(e_y) +V_{1,z}(e_y) V_{2,x}(e_z)$ \\
$+V_{1,x}(e_z) V_{2,y}(e_y) +V_{1,x}(e_y) V_{2,z}(e_z)$ \\
\midrule
$V_{1,x}(e_y) V_{2,y}(e_z) +V_{1,x}(e_z) V_{2,z}(e_y)$ \\
\midrule
$V_{1,y}(e_z) V_{2,x}(e_y) +V_{1,z}(e_y) V_{2,x}(e_z)$ \\
$-V_{1,x}(e_z) V_{2,y}(e_y) -V_{1,x}(e_y) V_{2,z}(e_z)$ \\
\midrule
$V_{1,z}(e_z) V_{2,x}(e_y) +V_{1,y}(e_y) V_{2,x}(e_z)$ \\
\midrule
$A_{1,x}(e_z) V_{2,x}(e_y) -A_{1,x}(e_y) V_{2,x}(e_z)$ \\
$-A_{1,y}(e_z) V_{2,y}(e_y) +A_{1,z}(e_y) V_{2,z}(e_z)$ \\
\midrule
$A_{1,z}(e_y) V_{2,y}(e_z) -A_{1,y}(e_z) V_{2,z}(e_y)$ \\
\midrule
$-A_{1,x}(e_z) V_{2,x}(e_y) +A_{1,x}(e_y) V_{2,x}(e_z)$ \\
$-A_{1,y}(e_z) V_{2,y}(e_y) +A_{1,z}(e_y) V_{2,z}(e_z)$ \\
\midrule
$A_{1,z}(e_z) V_{2,y}(e_y) -A_{1,y}(e_y) V_{2,z}(e_z)$ \\
\midrule
$A_{1,y}(e_y) V_{2,y}(e_z) -A_{1,z}(e_z) V_{2,z}(e_y)$ \\
\midrule
$A_{1,y}(e_z) A_{2,x}(e_y) +A_{1,z}(e_y) A_{2,x}(e_z)$ \\
$+A_{1,x}(e_z) A_{2,y}(e_y) +A_{1,x}(e_y) A_{2,z}(e_z)$ \\
\midrule
$A_{1,x}(e_y) A_{2,y}(e_z) +A_{1,x}(e_z) A_{2,z}(e_y)$ \\
\midrule
$A_{1,y}(e_z) A_{2,x}(e_y) +A_{1,z}(e_y) A_{2,x}(e_z)$ \\
$-A_{1,x}(e_z) A_{2,y}(e_y) -A_{1,x}(e_y) A_{2,z}(e_z)$ \\
\midrule
$A_{1,z}(e_z) A_{2,x}(e_y) +A_{1,y}(e_y) A_{2,x}(e_z)$ \\
\bottomrule
\end{tabular}
\addtolength{\tabcolsep}{-6pt}
\label{tab:mm-0nn-B2}
\end{table}

As the total momentum increases, the lattice symmetry is increasingly reduced. In particular, for momentum classes satisfying $|\vec{P}|^2 \geq 3$, the corresponding little groups become too small to provide meaningful symmetry classification, and the number of distinct operator structures grows large. Since $|\vec{P}|^2 \leq 2$ is typically sufficient for most spectroscopy calculations, we refrain from listing explicit operators beyond this range. Instead, we refer the reader to the \texttt{OpTion} package, which can generate operator bases for arbitrary momenta and symmetry classes as needed.

For the CM momenta class $\vec{P} = [n,m,p]$, where the little group reduces to the trivial group $C_1$, we do not provide an explicit list of operators. Instead, we note that without any protection of symmetry, all operators with particle momenta summing to $\vec{P}$ can contribute to the spectrum.

\section{List of baryon-baryon operators}
\label{sec:append_list_bb}
This section presents the list of distinguishable baryon–baryon operators, organized in Tab.~\ref{tab:bb-000}, \ref{tab:bb-00n}, \ref{tab:bb-0nn}, \ref{tab:bb-nnn}, \ref{tab:bb-nm0}, and \ref{tab:bb-nnm}. The associated operators are of manageable size, so we enumerate all relevant irreps except the trivial group.

\begin{table}[htbp]
\centering
\caption{Baryon-baryon operators of the cubic group $O_h$.}
\addtolength{\tabcolsep}{-5pt}
\begin{tabular}{cc}
\toprule
\multirow{4}{*}{$A_1^+$} & $N_{1,\frac{1}{2}}(0) N_{2,-\frac{1}{2}}(0) -N_{1,-\frac{1}{2}}(0) N_{2,\frac{1}{2}}(0)$ \\
\cmidrule(lr){2-2}
& $N_{1,\frac{1}{2}}(e_x) N_{2,-\frac{1}{2}}(-e_x) +N_{1,\frac{1}{2}}(-e_x) N_{2,-\frac{1}{2}}(e_x) +N_{1,\frac{1}{2}}(e_y) N_{2,-\frac{1}{2}}(-e_y) +N_{1,\frac{1}{2}}(-e_y) N_{2,-\frac{1}{2}}(e_y)$ \\
& $+N_{1,\frac{1}{2}}(e_z) N_{2,-\frac{1}{2}}(-e_z) +N_{1,\frac{1}{2}}(-e_z) N_{2,-\frac{1}{2}}(e_z) -N_{1,-\frac{1}{2}}(e_x) N_{2,\frac{1}{2}}(-e_x) -N_{1,-\frac{1}{2}}(-e_x) N_{2,\frac{1}{2}}(e_x)$ \\
& $-N_{1,-\frac{1}{2}}(e_y) N_{2,\frac{1}{2}}(-e_y) -N_{1,-\frac{1}{2}}(-e_y) N_{2,\frac{1}{2}}(e_y) -N_{1,-\frac{1}{2}}(e_z) N_{2,\frac{1}{2}}(-e_z) -N_{1,-\frac{1}{2}}(-e_z) N_{2,\frac{1}{2}}(e_z)$ \\
\midrule
\multirow{3}{*}{$A_1^-$} & $i N_{1,-\frac{1}{2}}(e_x) N_{2,-\frac{1}{2}}(-e_x) -i N_{1,-\frac{1}{2}}(-e_x) N_{2,-\frac{1}{2}}(e_x) +N_{1,-\frac{1}{2}}(e_y) N_{2,-\frac{1}{2}}(-e_y) -N_{1,-\frac{1}{2}}(-e_y) N_{2,-\frac{1}{2}}(e_y)$ \\
& $+i N_{1,\frac{1}{2}}(e_z) N_{2,-\frac{1}{2}}(-e_z) -i N_{1,\frac{1}{2}}(-e_z) N_{2,-\frac{1}{2}}(e_z) -i N_{1,\frac{1}{2}}(e_x) N_{2,\frac{1}{2}}(-e_x) +i N_{1,\frac{1}{2}}(-e_x) N_{2,\frac{1}{2}}(e_x)$ \\
& $+N_{1,\frac{1}{2}}(e_y) N_{2,\frac{1}{2}}(-e_y) -N_{1,\frac{1}{2}}(-e_y) N_{2,\frac{1}{2}}(e_y) +i N_{1,-\frac{1}{2}}(e_z) N_{2,\frac{1}{2}}(-e_z) -i N_{1,-\frac{1}{2}}(-e_z) N_{2,\frac{1}{2}}(e_z)$ \\
\midrule
\multirow{3}{*}{$E^+$} & $-N_{1,\frac{1}{2}}(e_x) N_{2,-\frac{1}{2}}(-e_x) -N_{1,\frac{1}{2}}(-e_x) N_{2,-\frac{1}{2}}(e_x) -N_{1,\frac{1}{2}}(e_y) N_{2,-\frac{1}{2}}(-e_y) -N_{1,\frac{1}{2}}(-e_y) N_{2,-\frac{1}{2}}(e_y)$ \\
\bottomrule
\end{tabular}
\addtolength{\tabcolsep}{5pt}
\label{tab:bb-000}
\end{table}

\begin{table}[htbp]
\centering
\caption{Baryon-baryon operators with $\vec{P} = [0,0,n]$.}
\addtolength{\tabcolsep}{6pt}
\begin{tabular}{cc}
\toprule
\multirow{2}{*}{$A_1$} & $N_{1,\frac{1}{2}}(0) N_{2,-\frac{1}{2}}(e_z) -N_{1,-\frac{1}{2}}(0) N_{2,\frac{1}{2}}(e_z)$ \\
\cmidrule(lr){2-2}
& $N_{1,\frac{1}{2}}(e_z) N_{2,-\frac{1}{2}}(0) -N_{1,-\frac{1}{2}}(e_z) N_{2,\frac{1}{2}}(0)$ \\
\midrule
\multirow{2}{*}{$A_2$} & $N_{1,\frac{1}{2}}(0) N_{2,-\frac{1}{2}}(e_z) +N_{1,-\frac{1}{2}}(0) N_{2,\frac{1}{2}}(e_z)$ \\
\cmidrule(lr){2-2}
& $N_{1,\frac{1}{2}}(e_z) N_{2,-\frac{1}{2}}(0) +N_{1,-\frac{1}{2}}(e_z) N_{2,\frac{1}{2}}(0)$ \\
\midrule
\multirow{2}{*}{$E$} & $N_{1,-\frac{1}{2}}(0) N_{2,-\frac{1}{2}}(e_z) -N_{1,\frac{1}{2}}(0) N_{2,\frac{1}{2}}(e_z)$ \\
\cmidrule(lr){2-2}
& $N_{1,-\frac{1}{2}}(e_z) N_{2,-\frac{1}{2}}(0) -N_{1,\frac{1}{2}}(e_z) N_{2,\frac{1}{2}}(0)$ \\
\bottomrule
\end{tabular}
\addtolength{\tabcolsep}{-6pt}
\label{tab:bb-00n}
\end{table}

\begin{table}[htbp]
\centering
\caption{Baryon-baryon operators with $\vec{P} = [0,n,n]$.}
\addtolength{\tabcolsep}{-5pt}
\begin{tabular}{cc}
\toprule
\multirow{2}{*}{$A_1$} & $N_{1,-\frac{1}{2}}(e_z) N_{2,-\frac{1}{2}}(e_y) -N_{1,-\frac{1}{2}}(e_y) N_{2,-\frac{1}{2}}(e_z) -N_{1,\frac{1}{2}}(e_z) N_{2,\frac{1}{2}}(e_y) +N_{1,\frac{1}{2}}(e_y) N_{2,\frac{1}{2}}(e_z)$ \\
\cmidrule(lr){2-2}
& $-N_{1,\frac{1}{2}}(e_z) N_{2,-\frac{1}{2}}(e_y) -N_{1,\frac{1}{2}}(e_y) N_{2,-\frac{1}{2}}(e_z) +N_{1,-\frac{1}{2}}(e_z) N_{2,\frac{1}{2}}(e_y) +N_{1,-\frac{1}{2}}(e_y) N_{2,\frac{1}{2}}(e_z)$ \\
\midrule
\multirow{2}{*}{$A_2$} & $N_{1,-\frac{1}{2}}(e_z) N_{2,-\frac{1}{2}}(e_y) +i N_{1,\frac{1}{2}}(e_y) N_{2,-\frac{1}{2}}(e_z) +N_{1,\frac{1}{2}}(e_z) N_{2,\frac{1}{2}}(e_y) +i N_{1,-\frac{1}{2}}(e_y) N_{2,\frac{1}{2}}(e_z)$ \\
\cmidrule(lr){2-2}
& $N_{1,\frac{1}{2}}(e_z) N_{2,-\frac{1}{2}}(e_y) -i N_{1,-\frac{1}{2}}(e_y) N_{2,-\frac{1}{2}}(e_z) +N_{1,-\frac{1}{2}}(e_z) N_{2,\frac{1}{2}}(e_y) -i N_{1,\frac{1}{2}}(e_y) N_{2,\frac{1}{2}}(e_z)$ \\
\midrule
\multirow{2}{*}{$B_1$} & $N_{1,-\frac{1}{2}}(e_z) N_{2,-\frac{1}{2}}(e_y) +N_{1,-\frac{1}{2}}(e_y) N_{2,-\frac{1}{2}}(e_z) -N_{1,\frac{1}{2}}(e_z) N_{2,\frac{1}{2}}(e_y) -N_{1,\frac{1}{2}}(e_y) N_{2,\frac{1}{2}}(e_z)$ \\
\cmidrule(lr){2-2}
& $-N_{1,\frac{1}{2}}(e_z) N_{2,-\frac{1}{2}}(e_y) +N_{1,\frac{1}{2}}(e_y) N_{2,-\frac{1}{2}}(e_z) +N_{1,-\frac{1}{2}}(e_z) N_{2,\frac{1}{2}}(e_y) -N_{1,-\frac{1}{2}}(e_y) N_{2,\frac{1}{2}}(e_z)$ \\
\midrule
\multirow{2}{*}{$B_2$} & $N_{1,-\frac{1}{2}}(e_z) N_{2,-\frac{1}{2}}(e_y) -i N_{1,\frac{1}{2}}(e_y) N_{2,-\frac{1}{2}}(e_z) +N_{1,\frac{1}{2}}(e_z) N_{2,\frac{1}{2}}(e_y) -i N_{1,-\frac{1}{2}}(e_y) N_{2,\frac{1}{2}}(e_z)$ \\
\cmidrule(lr){2-2}
& $N_{1,\frac{1}{2}}(e_z) N_{2,-\frac{1}{2}}(e_y) +i N_{1,-\frac{1}{2}}(e_y) N_{2,-\frac{1}{2}}(e_z) +N_{1,-\frac{1}{2}}(e_z) N_{2,\frac{1}{2}}(e_y) +i N_{1,\frac{1}{2}}(e_y) N_{2,\frac{1}{2}}(e_z)$ \\
\bottomrule
\end{tabular}
\addtolength{\tabcolsep}{5pt}
\label{tab:bb-0nn}
\end{table}

\begin{table}[htbp]
\centering
\scriptsize
\caption{Baryon-baryon operators with $\vec{P} = [n,n,n]$.}
\addtolength{\tabcolsep}{-5pt}
\begin{tabular}{cc}
\toprule
\multirow{13}{*}{$A_1$} & $N_{1,\frac{1}{2}}(0) N_{2,-\frac{1}{2}}(e_{x,y,z}) -N_{1,-\frac{1}{2}}(0) N_{2,\frac{1}{2}}(e_{x,y,z})$ \\
\cmidrule(lr){2-2}
& $(1+i) N_{1,-\frac{1}{2}}(e_z) N_{2,-\frac{1}{2}}(e_{x,y}) -N_{1,-\frac{1}{2}}(e_y) N_{2,-\frac{1}{2}}(e_{x,z}) +N_{1,\frac{1}{2}}(e_y) N_{2,-\frac{1}{2}}(e_{x,z}) -i N_{1,-\frac{1}{2}}(e_x) N_{2,-\frac{1}{2}}(e_{y,z})$ \\
& $-N_{1,\frac{1}{2}}(e_x) N_{2,-\frac{1}{2}}(e_{y,z}) -(1-i) N_{1,\frac{1}{2}}(e_z) N_{2,\frac{1}{2}}(e_{x,y}) +N_{1,-\frac{1}{2}}(e_y) N_{2,\frac{1}{2}}(e_{x,z}) +N_{1,\frac{1}{2}}(e_y) N_{2,\frac{1}{2}}(e_{x,z})$ \\
& $-N_{1,-\frac{1}{2}}(e_x) N_{2,\frac{1}{2}}(e_{y,z}) -i N_{1,\frac{1}{2}}(e_x) N_{2,\frac{1}{2}}(e_{y,z})$ \\
\cmidrule(lr){2-2}
& $-N_{1,\frac{1}{2}}(e_z) N_{2,-\frac{1}{2}}(e_{x,y}) -N_{1,\frac{1}{2}}(e_y) N_{2,-\frac{1}{2}}(e_{x,z}) -N_{1,\frac{1}{2}}(e_x) N_{2,-\frac{1}{2}}(e_{y,z}) +N_{1,-\frac{1}{2}}(e_z) N_{2,\frac{1}{2}}(e_{x,y})$ \\
& $+N_{1,-\frac{1}{2}}(e_y) N_{2,\frac{1}{2}}(e_{x,z}) +N_{1,-\frac{1}{2}}(e_x) N_{2,\frac{1}{2}}(e_{y,z})$ \\
\cmidrule(lr){2-2}
& $-i N_{1,-\frac{1}{2}}(e_{y,z}) N_{2,-\frac{1}{2}}(e_x) -N_{1,\frac{1}{2}}(e_{y,z}) N_{2,-\frac{1}{2}}(e_x) -N_{1,-\frac{1}{2}}(e_{x,z}) N_{2,-\frac{1}{2}}(e_y) +N_{1,\frac{1}{2}}(e_{x,z}) N_{2,-\frac{1}{2}}(e_y)$ \\
& $+(1+i) N_{1,-\frac{1}{2}}(e_{x,y}) N_{2,-\frac{1}{2}}(e_z) -N_{1,-\frac{1}{2}}(e_{y,z}) N_{2,\frac{1}{2}}(e_x) -i N_{1,\frac{1}{2}}(e_{y,z}) N_{2,\frac{1}{2}}(e_x) +N_{1,-\frac{1}{2}}(e_{x,z}) N_{2,\frac{1}{2}}(e_y)$ \\
\bottomrule
\end{tabular}
\addtolength{\tabcolsep}{5pt}
\label{tab:bb-nnn}
\end{table}

\begin{table}[htbp]
\centering
\caption{Baryon-baryon operators with $\vec{P} = [n,m,0]$.}
\addtolength{\tabcolsep}{6pt}
\begin{tabular}{cc}
\toprule
\multirow{6}{*}{$A$} & $N_{1,-\frac{1}{2}}(e_x) N_{2,\frac{1}{2}}(e_{x,y}) N_{1,\frac{1}{2}}(e_x) N_{2,-\frac{1}{2}}(e_{x,y})$ \\
\cmidrule(lr){2-2}
& $N_{1,-\frac{1}{2}}(e_{x,y}) N_{2,\frac{1}{2}}(e_x) N_{1,\frac{1}{2}}(e_{x,y}) N_{2,-\frac{1}{2}}(e_x)$ \\
\cmidrule(lr){2-2}
& $N_{1,-\frac{1}{2}}(e_y) N_{2,\frac{1}{2}}(e_{2x}) N_{1,\frac{1}{2}}(e_y) N_{2,-\frac{1}{2}}(e_{2x})$ \\
\cmidrule(lr){2-2}
& $-N_{1,-\frac{1}{2}}(e_{x,z}) N_{2,-\frac{1}{2}}(e_{x,y,-z}) +N_{1,-\frac{1}{2}}(e_{x,-z}) N_{2,-\frac{1}{2}}(e_{x,y,z})$ \\
\cmidrule(lr){2-2}
& $N_{1,-\frac{1}{2}}(e_{x,z}) N_{2,\frac{1}{2}}(e_{x,y,-z}) +N_{1,-\frac{1}{2}}(e_{x,-z}) N_{2,\frac{1}{2}}(e_{x,y,z})$ \\
\cmidrule(lr){2-2}
& $N_{1,\frac{1}{2}}(e_{x,z}) N_{2,-\frac{1}{2}}(e_{x,y,-z}) +N_{1,\frac{1}{2}}(e_{x,-z}) N_{2,-\frac{1}{2}}(e_{x,y,z})$ \\
\midrule
\multirow{6}{*}{$B$} & $N_{1,-\frac{1}{2}}(e_x) N_{2,-\frac{1}{2}}(e_{x,y}) N_{1,\frac{1}{2}}(e_x) N_{2,\frac{1}{2}}(e_{x,y})$ \\
\cmidrule(lr){2-2}
& $N_{1,-\frac{1}{2}}(e_{x,y}) N_{2,-\frac{1}{2}}(e_x) N_{1,\frac{1}{2}}(e_{x,y}) N_{2,\frac{1}{2}}(e_x)$ \\
\cmidrule(lr){2-2}
& $N_{1,-\frac{1}{2}}(e_y) N_{2,-\frac{1}{2}}(e_{2x}) N_{1,\frac{1}{2}}(e_y) N_{2,\frac{1}{2}}(e_{2x})$ \\
\cmidrule(lr){2-2}
& $N_{1,-\frac{1}{2}}(e_{x,z}) N_{2,-\frac{1}{2}}(e_{x,y,-z}) +N_{1,-\frac{1}{2}}(e_{x,-z}) N_{2,-\frac{1}{2}}(e_{x,y,z})$ \\
\cmidrule(lr){2-2}
& $-N_{1,-\frac{1}{2}}(e_{x,z}) N_{2,\frac{1}{2}}(e_{x,y,-z}) +N_{1,-\frac{1}{2}}(e_{x,-z}) N_{2,\frac{1}{2}}(e_{x,y,z})$ \\
\cmidrule(lr){2-2}
& $-N_{1,\frac{1}{2}}(e_{x,z}) N_{2,-\frac{1}{2}}(e_{x,y,-z}) +N_{1,\frac{1}{2}}(e_{x,-z}) N_{2,-\frac{1}{2}}(e_{x,y,z})$ \\
\bottomrule
\end{tabular}
\addtolength{\tabcolsep}{-6pt}
\label{tab:bb-nm0}
\end{table}

\begin{table}[htbp]
\centering
\caption{Baryon-baryon operators with $\vec{P} = [n,n,m]$.}
\addtolength{\tabcolsep}{6pt}
\begin{tabular}{cc}
\toprule
\multirow{8}{*}{$A$} & $N_{1,-\frac{1}{2}}(e_z) N_{2,-\frac{1}{2}}(e_{x,y,z}) +i N_{1,\frac{1}{2}}(e_z) N_{2,\frac{1}{2}}(e_{x,y,z})$ \\
\cmidrule(lr){2-2}
& $-N_{1,\frac{1}{2}}(e_z) N_{2,-\frac{1}{2}}(e_{x,y,z}) +N_{1,-\frac{1}{2}}(e_z) N_{2,\frac{1}{2}}(e_{x,y,z})$ \\
\cmidrule(lr){2-2}
& $N_{1,-\frac{1}{2}}(e_{y,z}) N_{2,-\frac{1}{2}}(e_{x,z}) +i N_{1,\frac{1}{2}}(e_{x,z}) N_{2,\frac{1}{2}}(e_{y,z})$ \\
\cmidrule(lr){2-2}
& $-N_{1,\frac{1}{2}}(e_{x,z}) N_{2,-\frac{1}{2}}(e_{y,z}) +N_{1,-\frac{1}{2}}(e_{y,z}) N_{2,\frac{1}{2}}(e_{x,z})$ \\
\cmidrule(lr){2-2}
& $N_{1,\frac{1}{2}}(e_{y,z}) N_{2,-\frac{1}{2}}(e_{x,z}) -N_{1,-\frac{1}{2}}(e_{x,z}) N_{2,\frac{1}{2}}(e_{y,z})$ \\
\cmidrule(lr){2-2}
& $-i N_{1,-\frac{1}{2}}(e_{x,z}) N_{2,-\frac{1}{2}}(e_{y,z}) +N_{1,\frac{1}{2}}(e_{y,z}) N_{2,\frac{1}{2}}(e_{x,z})$ \\
\cmidrule(lr){2-2}
& $N_{1,-\frac{1}{2}}(e_{x,y,z}) N_{2,-\frac{1}{2}}(e_z) +i N_{1,\frac{1}{2}}(e_{x,y,z}) N_{2,\frac{1}{2}}(e_z)$ \\
\cmidrule(lr){2-2}
& $-N_{1,\frac{1}{2}}(e_{x,y,z}) N_{2,-\frac{1}{2}}(e_z) +N_{1,-\frac{1}{2}}(e_{x,y,z}) N_{2,\frac{1}{2}}(e_z)$ \\
\bottomrule
\end{tabular}
\addtolength{\tabcolsep}{-6pt}
\label{tab:bb-nnm}
\end{table}

\section{List of meson-baryon operators}
\label{sec:append_list_mb}
This section presents meson–baryon operators, organized in Tab.~\ref{tab:mb-000-G1}, \ref{tab:mb-000-G2}, \ref{tab:mb-000-H+}, \ref{tab:mb-000-H-}, \ref{tab:mb-00n}, and \ref{tab:mb-0nn}. Since the total spin in these systems is half-integer, the operators live in the double cover of the relevant lattice symmetry group.

\begin{table}[htbp]
\centering
\scriptsize
\caption{Meson-baryon operators in $G_1^{\pm}$ irrep of the double cover cubic group $O_h^D$.}
\addtolength{\tabcolsep}{-5pt}
\begin{tabular}{cc}
\toprule
\multirow{22}{*}{$G_1^+$} & $P_{1}(e_x) N_{2,-\frac{1}{2}}(-e_x) -P_{1}(-e_x) N_{2,-\frac{1}{2}}(e_x) -i P_{1}(e_y) N_{2,-\frac{1}{2}}(-e_y) +i P_{1}(-e_y) N_{2,-\frac{1}{2}}(e_y)$ \\
& $+P_{1}(e_z) N_{2,\frac{1}{2}}(-e_z) -P_{1}(-e_z) N_{2,\frac{1}{2}}(e_z)$ \\
\cmidrule(lr){2-2}
& $S_{1}(0) N_{2,\frac{1}{2}}(0)$ \\
\cmidrule(lr){2-2}
& $S_{1}(e_x) N_{2,\frac{1}{2}}(-e_x) +S_{1}(-e_x) N_{2,\frac{1}{2}}(e_x) +S_{1}(e_y) N_{2,\frac{1}{2}}(-e_y) +S_{1}(-e_y) N_{2,\frac{1}{2}}(e_y)$ \\
& $+S_{1}(e_z) N_{2,\frac{1}{2}}(-e_z) +S_{1}(-e_z) N_{2,\frac{1}{2}}(e_z)$ \\
\cmidrule(lr){2-2}
& $2 N_{2,\frac{1}{2}}(e_x) V_{1,x}(-e_x) -2 N_{2,\frac{1}{2}}(-e_x) V_{1,x}(e_x) -i N_{2,\frac{1}{2}}(e_y) V_{1,x}(-e_y) +i N_{2,\frac{1}{2}}(-e_y) V_{1,x}(e_y)$ \\
& $+N_{2,-\frac{1}{2}}(e_z) V_{1,x}(-e_z) -N_{2,-\frac{1}{2}}(-e_z) V_{1,x}(e_z) +i N_{2,\frac{1}{2}}(e_x) V_{1,y}(-e_x) -i N_{2,\frac{1}{2}}(-e_x) V_{1,y}(e_x)$ \\
& $+2 N_{2,\frac{1}{2}}(e_y) V_{1,y}(-e_y) -2 N_{2,\frac{1}{2}}(-e_y) V_{1,y}(e_y) -i N_{2,-\frac{1}{2}}(e_z) V_{1,y}(-e_z) +i N_{2,-\frac{1}{2}}(-e_z) V_{1,y}(e_z)$ \\
\bottomrule
\end{tabular}
\addtolength{\tabcolsep}{5pt}
\label{tab:mb-000-G1}
\end{table}

\begin{table}[htbp]
\centering
\scriptsize
\caption{Meson-baryon operators in $G_2^{\pm}$ irrep of the double cover cubic group $O_h^D$.}
\addtolength{\tabcolsep}{-5pt}
\begin{tabular}{cc}
\toprule
\multirow{8}{*}{$G_2^+$} & $i N_{2,\frac{1}{2}}(e_y) V_{1,x}(-e_y) -i N_{2,\frac{1}{2}}(-e_y) V_{1,x}(e_y) +N_{2,-\frac{1}{2}}(e_z) V_{1,x}(-e_z) -N_{2,-\frac{1}{2}}(-e_z) V_{1,x}(e_z)$ \\
& $+i N_{2,\frac{1}{2}}(e_x) V_{1,y}(-e_x) -i N_{2,\frac{1}{2}}(-e_x) V_{1,y}(e_x) +i N_{2,-\frac{1}{2}}(e_z) V_{1,y}(-e_z) -i N_{2,-\frac{1}{2}}(-e_z) V_{1,y}(e_z)$ \\
& $+N_{2,-\frac{1}{2}}(e_x) V_{1,z}(-e_x) -N_{2,-\frac{1}{2}}(-e_x) V_{1,z}(e_x) +i N_{2,-\frac{1}{2}}(e_y) V_{1,z}(-e_y) -i N_{2,-\frac{1}{2}}(-e_y) V_{1,z}(e_y)$ \\
\cmidrule(lr){2-2}
& $i A_{1,y}(e_x) N_{2,-\frac{1}{2}}(-e_x) +i A_{1,y}(-e_x) N_{2,-\frac{1}{2}}(e_x) +A_{1,x}(e_y) N_{2,-\frac{1}{2}}(-e_y) +A_{1,x}(-e_y) N_{2,-\frac{1}{2}}(e_y)$ \\
& $-A_{1,x}(e_z) N_{2,-\frac{1}{2}}(-e_z) -i A_{1,y}(e_z) N_{2,-\frac{1}{2}}(-e_z) -A_{1,x}(-e_z) N_{2,-\frac{1}{2}}(e_z) -i A_{1,y}(-e_z) N_{2,-\frac{1}{2}}(e_z)$ \\
& $+A_{1,z}(e_x) N_{2,\frac{1}{2}}(-e_x) +A_{1,z}(-e_x) N_{2,\frac{1}{2}}(e_x) -A_{1,z}(e_y) N_{2,\frac{1}{2}}(-e_y) -A_{1,z}(-e_y) N_{2,\frac{1}{2}}(e_y)$ \\
\midrule
\multirow{8}{*}{$G_2^-$} & $N_{2,-\frac{1}{2}}(e_y) V_{1,x}(-e_y) + N_{2,-\frac{1}{2}}(-e_y) V_{1,x}(e_y) - N_{2,-\frac{1}{2}}(e_z) V_{1,x}(-e_z) - N_{2,-\frac{1}{2}}(-e_z) V_{1,x}(e_z)$ \\
& $+ i N_{2,-\frac{1}{2}}(e_x) V_{1,y}(-e_x) + i N_{2,-\frac{1}{2}}(-e_x) V_{1,y}(e_x) - i N_{2,-\frac{1}{2}}(e_z) V_{1,y}(-e_z) - i N_{2,-\frac{1}{2}}(-e_z) V_{1,y}(e_z)$ \\
\bottomrule
\end{tabular}
\addtolength{\tabcolsep}{5pt}
\label{tab:mb-000-G2}
\end{table}

\begin{table}[htbp]
\centering
\small
\caption{Meson-baryon operators in $H^+$ irrep of the double cover cubic group $O_h^D$.}
\addtolength{\tabcolsep}{-5pt}
\begin{tabular}{cc}
\toprule
\multirow{25}{*}{$H^+$} & $P_{1}[e_x] N_{2,\frac{1}{2}}[-e_x] - P_{1}[-e_x] N_{2,\frac{1}{2}}[e_x] - i P_{1}[e_y] N_{2,\frac{1}{2}}[-e_y] + i P_{1}[-e_y] N_{2,\frac{1}{2}}[e_y]$ \\
\cmidrule(lr){2-2}
& $S_{1}[e_x] N_{2,-\frac{1}{2}}[-e_x] + S_{1}[-e_x] N_{2,-\frac{1}{2}}[e_x] - S_{1}[e_y] N_{2,-\frac{1}{2}}[-e_y] - S_{1}[-e_y] N_{2,-\frac{1}{2}}[e_y]$ \\
\cmidrule(lr){2-2}
& $3 N_{2,-\frac{1}{2}}[e_x] V_{1,x}[-e_x] - 3 N_{2,-\frac{1}{2}}[-e_x] V_{1,x}[e_x] + 2 i N_{2,-\frac{1}{2}}[e_y] V_{1,x}[-e_y] - 2 i N_{2,-\frac{1}{2}}[-e_y] V_{1,x}[e_y]$ \\
& $- N_{2,\frac{1}{2}}[e_z] V_{1,x}[-e_z] + N_{2,\frac{1}{2}}[-e_z] V_{1,x}[e_z] + 2 i N_{2,-\frac{1}{2}}[e_x] V_{1,y}[-e_x] - 2 i N_{2,-\frac{1}{2}}[-e_x] V_{1,y}[e_x]$ \\
& $- 3 N_{2,-\frac{1}{2}}[e_y] V_{1,y}[-e_y] + 3 N_{2,-\frac{1}{2}}[-e_y] V_{1,y}[e_y] + i N_{2,\frac{1}{2}}[e_z] V_{1,y}[-e_z] - i N_{2,\frac{1}{2}}[-e_z] V_{1,y}[e_z]$ \\
& $- N_{2,\frac{1}{2}}[e_x] V_{1,z}[-e_x] + N_{2,\frac{1}{2}}[-e_x] V_{1,z}[e_x] + i N_{2,\frac{1}{2}}[e_y] V_{1,z}[-e_y] - i N_{2,\frac{1}{2}}[-e_y] V_{1,z}[e_y]$ \\
\cmidrule(lr){2-2}
& $i N_{2,-\frac{1}{2}}[e_y] V_{1,x}[-e_y] - i N_{2,-\frac{1}{2}}[-e_y] V_{1,x}[e_y] + N_{2,\frac{1}{2}}[e_z] V_{1,x}[-e_z] - N_{2,\frac{1}{2}}[-e_z] V_{1,x}[e_z]$ \\
& $+ i N_{2,-\frac{1}{2}}[e_x] V_{1,y}[-e_x] - i N_{2,-\frac{1}{2}}[-e_x] V_{1,y}[e_x] - i N_{2,\frac{1}{2}}[e_z] V_{1,y}[-e_z] + i N_{2,\frac{1}{2}}[-e_z] V_{1,y}[e_z]$ \\
\bottomrule
\end{tabular}
\addtolength{\tabcolsep}{5pt}
\label{tab:mb-000-H+}
\end{table}

\begin{table}[htbp]
\centering
\small
\caption{Meson-baryon operators in $H^-$ irrep of the double cover cubic group $O_h^D$.}
\addtolength{\tabcolsep}{-5pt}
\begin{tabular}{cc}
\toprule
\multirow{25}{*}{$H^-$} & $P_1[e_x] N_{2,-\frac{1}{2}}[-e_x] + P_1[-e_x] N_{2,-\frac{1}{2}}[e_x] - P_1[e_y] N_{2,-\frac{1}{2}}[-e_y] - P_1[-e_y] N_{2,-\frac{1}{2}}[e_y]$ \\
\cmidrule(lr){2-2}
& $S_1[e_x] N_{2,\frac{1}{2}}[-e_x] - S_1[-e_x] N_{2,\frac{1}{2}}[e_x] - i S_1[e_y] N_{2,\frac{1}{2}}[-e_y] + i S_1[-e_y] N_{2,\frac{1}{2}}[e_y]$ \\
\cmidrule(lr){2-2}
& $i N_{2,\frac{1}{2}}[0] V_{1,x}[0] + N_{2,\frac{1}{2}}[0] V_{1,y}[0]$ \\
\cmidrule(lr){2-2}
& $3 N_{2,\frac{1}{2}}[e_x] V_{1,x}[-e_x] + 3 N_{2,\frac{1}{2}}[-e_x] V_{1,x}[e_x] - 2 N_{2,\frac{1}{2}}[e_y] V_{1,x}[-e_y] - 2 N_{2,\frac{1}{2}}[-e_y] V_{1,x}[e_y]$ \\
& $- N_{2,\frac{1}{2}}[e_z] V_{1,x}[-e_z] - N_{2,\frac{1}{2}}[-e_z] V_{1,x}[e_z] + 2 i N_{2,\frac{1}{2}}[e_x] V_{1,y}[-e_x] + 2 i N_{2,\frac{1}{2}}[-e_x] V_{1,y}[e_x]$ \\
& $- 3 i N_{2,\frac{1}{2}}[e_y] V_{1,y}[-e_y] - 3 i N_{2,\frac{1}{2}}[-e_y] V_{1,y}[e_y] + i N_{2,\frac{1}{2}}[e_z] V_{1,y}[-e_z] + i N_{2,\frac{1}{2}}[-e_z] V_{1,y}[e_z]$ \\
& $- N_{2,-\frac{1}{2}}[e_x] V_{1,z}[-e_x] - N_{2,-\frac{1}{2}}[-e_x] V_{1,z}[e_x] + N_{2,-\frac{1}{2}}[e_y] V_{1,z}[-e_y] + N_{2,-\frac{1}{2}}[-e_y] V_{1,z}[e_y]$ \\
\cmidrule(lr){2-2}
& $N_{2,\frac{1}{2}}[e_y] V_{1,x}[-e_y] + N_{2,\frac{1}{2}}[-e_y] V_{1,x}[e_y] - N_{2,\frac{1}{2}}[e_z] V_{1,x}[-e_z] - N_{2,\frac{1}{2}}[-e_z] V_{1,x}[e_z]$ \\
\bottomrule
\end{tabular}
\addtolength{\tabcolsep}{5pt}
\label{tab:mb-000-H-}
\end{table}

\begin{table}[htbp]
\centering
\caption{Meson-baryon operators with $\vec{P} = [0,0,n]$.}
\addtolength{\tabcolsep}{6pt}
\begin{tabular}{cc}
\toprule
\multirow{12}{*}{$G_1$} & $P_{1}(0) N_{2,\frac{1}{2}}(e_z)$ \\
\cmidrule(lr){2-2}
& $P_{1}(e_z) N_{2,\frac{1}{2}}(0)$ \\
\cmidrule(lr){2-2}
& $S_{1}(0) N_{2,\frac{1}{2}}(e_z)$ \\
\cmidrule(lr){2-2}
& $S_{1}(e_z) N_{2,\frac{1}{2}}(0)$ \\
\cmidrule(lr){2-2}
& $N_{2,\frac{1}{2}}(e_z) V_{1,z}(0)$ \\
\cmidrule(lr){2-2}
& $N_{2,-\frac{1}{2}}(e_z) V_{1,x}(0) -i N_{2,-\frac{1}{2}}(e_z) V_{1,y}(0)$ \\
\cmidrule(lr){2-2}
& $N_{2,\frac{1}{2}}(0) V_{1,z}(e_z)$ \\
\cmidrule(lr){2-2}
& $N_{2,-\frac{1}{2}}(0) V_{1,x}(e_z) -i N_{2,-\frac{1}{2}}(0) V_{1,y}(e_z)$ \\
\cmidrule(lr){2-2}
& $A_{1,z}(0) N_{2,\frac{1}{2}}(e_z)$ \\
\cmidrule(lr){2-2}
& $A_{1,x}(0) N_{2,-\frac{1}{2}}(e_z) -i A_{1,y}(0) N_{2,-\frac{1}{2}}(e_z)$ \\
\cmidrule(lr){2-2}
& $A_{1,z}(e_z) N_{2,\frac{1}{2}}(0)$ \\
\cmidrule(lr){2-2}
& $A_{1,x}(e_z) N_{2,-\frac{1}{2}}(0) -i A_{1,y}(e_z) N_{2,-\frac{1}{2}}(0)$ \\
\midrule
\multirow{4}{*}{$G_2$} & $N_{2,-\frac{1}{2}}(e_z) V_{1,x}(0) +i N_{2,-\frac{1}{2}}(e_z) V_{1,y}(0)$ \\
\cmidrule(lr){2-2}
& $N_{2,-\frac{1}{2}}(0) V_{1,x}(e_z) +i N_{2,-\frac{1}{2}}(0) V_{1,y}(e_z)$ \\
\cmidrule(lr){2-2}
& $A_{1,x}(0) N_{2,-\frac{1}{2}}(e_z) +i A_{1,y}(0) N_{2,-\frac{1}{2}}(e_z)$ \\
\cmidrule(lr){2-2}
& $A_{1,x}(e_z) N_{2,-\frac{1}{2}}(0) +i A_{1,y}(e_z) N_{2,-\frac{1}{2}}(0)$ \\
\bottomrule
\end{tabular}
\addtolength{\tabcolsep}{-6pt}
\label{tab:mb-00n}
\end{table}

\begin{table}[htbp]
\centering
\caption{Meson-baryon operators with $\vec{P} = [0,n,n]$.}
\addtolength{\tabcolsep}{6pt}
\begin{tabular}{cc}
\toprule
\multirow{16}{*}{$G$} & $P_{1}(e_z) N_{2,-\frac{1}{2}}(e_y) +i P_{1}(e_y) N_{2,\frac{1}{2}}(e_z)$ \\
\cmidrule(lr){2-2}
& $-i P_{1}(e_y) N_{2,-\frac{1}{2}}(e_z) +P_{1}(e_z) N_{2,\frac{1}{2}}(e_y)$ \\
\cmidrule(lr){2-2}
& $S_{1}(e_z) N_{2,-\frac{1}{2}}(e_y) -S_{1}(e_y) N_{2,-\frac{1}{2}}(e_z)$ \\
\cmidrule(lr){2-2}
& $S_{1}(e_z) N_{2,\frac{1}{2}}(e_y) +S_{1}(e_y) N_{2,\frac{1}{2}}(e_z)$ \\
\cmidrule(lr){2-2}
& $-i N_{2,\frac{1}{2}}(e_z) V_{1,x}(e_y) +N_{2,-\frac{1}{2}}(e_y) V_{1,x}(e_z) +i N_{2,-\frac{1}{2}}(e_y) V_{1,y}(e_z) -i N_{2,-\frac{1}{2}}(e_z) V_{1,z}(e_y)$ \\
\cmidrule(lr){2-2}
& $i N_{2,-\frac{1}{2}}(e_z) V_{1,x}(e_y) +N_{2,\frac{1}{2}}(e_y) V_{1,x}(e_z) +i N_{2,\frac{1}{2}}(e_y) V_{1,y}(e_z) +i N_{2,\frac{1}{2}}(e_z) V_{1,z}(e_y)$ \\
\cmidrule(lr){2-2}
& $N_{2,-\frac{1}{2}}(e_z) V_{1,y}(e_y) -N_{2,-\frac{1}{2}}(e_y) V_{1,z}(e_z)$ \\
\cmidrule(lr){2-2}
& $N_{2,\frac{1}{2}}(e_z) V_{1,y}(e_y) +N_{2,\frac{1}{2}}(e_y) V_{1,z}(e_z)$ \\
\bottomrule
\end{tabular}
\addtolength{\tabcolsep}{-6pt}
\label{tab:mb-0nn}
\end{table}

\clearpage

\bibliographystyle{JHEP}      
\bibliography{paper}

\end{document}